\documentclass[12pt]{iopart}
\usepackage{graphicx}
\usepackage{color}
\begin{document}

\title[Moody's Correlated Binomial Default Distribution]
{
Moody's Correlated Binomial Default Distributions for 
Inhomogeneous Portfolios
}

\author{S.Mori \dag 
\footnote[1]{mori@sci.kitasato-u.ac.jp}, 
K. Kitsukawa \ddag \footnote[2]{
kenji.kitsukawa@daiwasmbc.co.jp}
and M. Hisakado 
\P \footnote[3]{
masato\_hisakado@standardandpoors.com
}    
}
 
\address{\dag\ Department of Physics, School of Science,
Kitasato University, Kitasato 1-15-1 , Sagamihara, Kanagawa 228-8555, Japan}

\address{\P\  
Daiwa Securities SMBC,  
Marunouchi 1-9-1, Chiyoda-ku, Tokyo 100-6753, Japan}

\address{\ddag\ 
Standard \& Poor's, Marunouchi 1-6-5, Chiyoda-ku, Tokyo 100-0005, Japan}

\vspace*{4.5cm}

\begin{abstract}
This paper generalizes Moody's 
correlated binomial default distribution for homogeneous (exchangeable)  
 credit portfolio, which is introduced by 
Witt, to the case of 
inhomogeneous portfolios.
As inhomogeneous portfolios, we consider two cases.
In the first case, we  treat a portfolio whose assets have 
uniform default correlation and non-uniform 
default probabilities. We obtain the default probability
distribution and study the effect of the 
inhomogeneity on it. 
The second case corresponds to a portfolio with inhomogeneous 
default correlation. Assets are categorized in several different 
sectors and  the inter-sector and intra-sector correlations are not the
 same. We construct the joint default probabilities and 
 obtain the default probability distribution. We show that as the
number of assets in each sector decreases, inter-sector
correlation becomes more important than  intra-sector correlation. 
We study the maximum values of the inter-sector
default correlation.
Our generalization method can 
be applied to any correlated binomial default distribution model which has
explicit relations to  the conditional default probabilities or conditional
default correlations, e.g. Credit Risk${}^{+}$, 
 implied default distributions. 
We also compare 
some popular CDO pricing models from the viewpoint
of the range of the implied tranche correlation. 
\end{abstract}

\submitto{Quantitative Finance: Revised}

\maketitle

\section{Introduction}

The modeling of portfolio
credit risk and default correlation  are hot topics and pose entirely
new problems \cite{Fabozzi,Schonbucher,Duffie2,Hull,Finger2}. 
CDOs are
financial innovations to securitize portfolios of defaultable assets.
 Many probabilistic models 
have been studied in order to price CDO tranches 
\cite{Cifuentes,Martin,
Finger,Duffie,Li,Vasicek,Schonbucher2,Duffie2,Andersen,Davis,Zhou}.
Most of them are implemented with Monte Carlo simulations and
as the number of names in a  portfolio increases, the computational
time increases. The Factor approach uses a small number of latent
factors that induce the default dependency \cite{Vasicek}. 
Conditionally on the 
latent variables values, default events are independent. It becomes easy
to calculate the loss (default) distribution function. Along this line,
some semi-explicit  expressions of most relevant quantities 
were obtained \cite{Laurent}.

On the other hand, correlated binomial models were also 
proposed to describe the default dependency structures.
 The first one is a one-factor exchangeable version of CreditRisk
${}^{+}$ \cite{CR,Frey1,Frey2}. The aggregate loss distribution function 
is  given by the beta-binomial distribution (BBD).
 The second one is  Moody's correlated binomial default distribution
 model, which was introduced by Witt \cite{Witt}.( hereafter the
MCB model)  
The authors also consider the applicability of the long-range Ising
model \cite{Molins,Mori}. 
These  models use Bernoulli  random variables.
Differences stem  from different definitions of the
conditional correlations \cite{Hisakado}. 
In the MCB model \cite{Witt}, 
the  conditional  default correlation 
between assets is set to be constant
irrespective  of the number of defaults. 
Those of BBD decay with an increase in default.
We are able to adapt a suitable form for the conditional correlations. 
Recently 
it has become possible to
calibrate $\rho_{n}$ from implied default distributions \cite{Mori2}. 
 By using the ``implied correlated'' binomial model,
 whose conditional correlations are those of the implied distribution,   
 it may become easy to estimate 
hedge ratios and so forth.

The advantage of 
these  correlated binomial  
models come from  the fact  that
they are easier to evaluate than other more refined models.
If a probabilistic model is  implemented by a  Monte Carlo 
simulation, 
the evaluation of the price of these derivatives consumes much computer
time and the inverse process to obtain the model parameters becomes
tedious work. With the above correlated binomial
 models, one can estimate 
the model parameters from the CDO  
premiums more easily. 
 However, these models
are formulated only for homogeneous portfolios,  
where the assets  are 
exchangeable  and they have the same default
probability $p$ and default correlation $\rho$. 
Generalization to  more realistic  inhomogeneous 
portfolios where assets have different default probabilities 
 and different default correlations  should be done.

In this paper, we show how to generalize  Moody's correlated
binomial default distribution (MCB) model to two types of 
inhomogeneous portfolios. 
Our generalization method can be applied 
to other correlated binomial models,  including implied 
correlated binomial distributions, 
by changing the condition on $\rho_{n}$.
We obtain the default probability function
$P_{N}(n)$ and examine the dependence of the expected loss
rates of the tranches on the inhomogeneities.
With the proposed model, we also estimate the implied 
values of the default correlation. 
 Comparison of the range of the tranche
correlations with those of the Gaussian copula model and  BBD model 
are also performed.

About the organization in  this paper, we start with a short  review of 
 Moody's correlated binomial default distribution (MCB) model
in \Sref{MCB}. 
The dependence of  $\rho_n$ 
on the number of defaults $n$
is compared with the BBD and Gaussian copula. 
\Sref{model} is the main part of the
paper. We show how to couple multiple MCB models as a portfolios 
credit risk model for inhomogeneous portfolio. 
In the first subsection, we couple two Bernoulli 
 random 
variables $X,Y$ 
and recall on the limit of the correlation 
$\rho_{xy}$ between them. 
In the next subsection, 
we couple an MCB model of  $N$ assets $X_{1},X_{2},\cdots, X_{N}$ 
with a random variable $Y$ and we study 
 the maximum value of the correlation between 
$X_{i}$ and $Y$.  
Then we couple two MCB models with $N$ and $M$ assets. 
Choosing the model parameters properly, 
we construct an MCB model for an inhomogeneous
default probability case and obtain the default distribution 
function. 
The last subsection is devoted  to an inhomogeneous default correlation
case. Assets are categorized in different sectors and
 inter-sector
and intra-sector default correlations are not the same.  
We consider a portfolio with $K$ sectors  
and $k-$th sector contains
$N_{k}$ assets. Within each sector, the portfolio is homogeneous
and it has parameters as $p_{k}$ and $\rho_{k}$.
The inter-sector default
correlations are not the same and they depend on the choice of sector
pairs. We construct the joint default probabilities and the
default  probability function $P_{N}(n)$
for the portfolio explicitly. 
In \Sref{implied_c}, using the above results, 
we estimate the implied 
default correlation for each tranche from a 
CDO's market quotes ( iTraxx-CJ Series 2).
We compare the range of the correlations of MCBs, BBD and the Gaussian
copula model.  
We conclude with some remarks and future problems.

\section{Moody's correlated binomial default distribution}
\label{MCB}

\begin{figure}[htbp]
 \begin{center}
  \includegraphics[width=4.0cm]{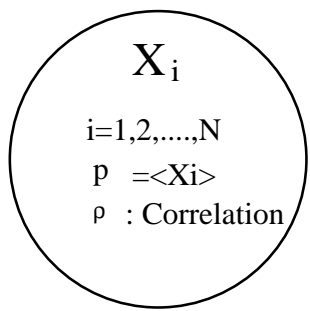}
  \caption{Homogeneous portfolio with $N$ assets.
The assets are exchangeable and  
the default probability is $p$ and the
  default correlation is $\rho.$ 
 The state of i-th asset is described by 
a Bernoulli random variable $X_{i}.$ }
  \label{single}
 \end{center}
\end{figure}

We review the definitions and some properties of Moody's correlated binomial
default distribution (MCB) model. 
We consider a homogeneous portfolio, which is
composed of 
exchangeable  
$N$ assets. Here the term ``homogeneous'' means that 
 the constituent  
assets are exchangeable and 
their default 
probabilities and default correlations 
are uniform.  We denote
them as $p$ and $\rho$.  Bernoulli  
random variables $X_{i}$ show the
states of the $i$-th assets. $X_{i}=1$ means that the asset is defaulted
and the non-default state is represented as $X_{i}=0$. 
The joint default probabilities are denoted as 
\begin{equation}
P(x_{1},x_{2},\cdots,x_{N})=\mbox{Prob}(X_{1}=x_{1},X_{2}=x_{2},
\cdots,X_{N}=x_{N}).
\end{equation}
In order to determine $P(x_{1},x_{2},\cdots,x_{N})$, we need $2^{N}-1$
conditions for them. Here $2^{N}$ corresponds to the number of possible 
configurations and $-1$ comes from the overall normalization condition for the 
joint probabilities.
From the assumption of the homogeneity for the
portfolio, the number of degrees of freedom of the 
joint probabilities are reduced. The probability 
for $n$ defaults and $N-n$ non-defaults is the  same for any configuration
$(x_{1},x_{2},\cdots,x_{N})$ with $\sum_{i=1}^{N}x_{i}=n$.
The number of defaults $n$ is ranged from $0$ to $N$ 
and considering the overall
normalization condition, remaining degrees of freedom are $N$. 

In the MCB model, the conditional default probabilities are
introduced. We denote $p_{n}$ as the default probability for any assets 
under the condition that any other $n$ assets of the portfolio are
defaulted. To exemplify the situation concretely, we take the $n$ assets 
as the first $n$ of $N$ assets, we denote them with $n^{'}$ as $n^{'}
=1,2,\cdots,n$. The condition that they are defaulted is written
concisely as $\prod_{n^{'}=1}^{n}X_{n^{'}}=1$. The conditional 
default probability for $n+1$-th assets under the condition of  $n$
defaults can then be written as
\begin{equation}
p_{n}=<X_{n+1}|\prod_{n^{'}=1}^{n}X_{n^{'}}=1>.
\end{equation}
Here, $<A | C >$ means the expected  value of random variable $A$
under condition $C$ is satisfied.
$X_{n}$ takes $1$ for $n$-th asset default, 
$<X_{n}| C>$ corresponds to its default probability under condition $C$. 
Of course, any asset from $k=n+1,\cdots,N$ can be 
chosen in the evaluation
of the expected value for $p_{n}$ under the condition that 
$\prod_{n^{'}=1}^{n}X_{n^{'}}=1$.
$p_{0}$ is nothing but the default probability $p$.

$N$ independent conditional default probabilities
$p_{n} (i=0,\cdots,N-1)$ are determined by the following condition on
the default correlations.
\begin{equation}
\mbox{Cor}(X_{n+1},X_{n+2}|\prod_{n^{'}=1}^{n}X_{n^{'}}=1)=\rho .
\label{corr_MCB}
\end{equation}
Here, $\mbox{Cor}(X,Y | C)$ is defined as
\begin{equation}
\mbox{Cor}(X,Y | C)=\frac{
<XY |C>-<X|C><Y|C>}{\sqrt{<X|C>(1-<X|C>)<Y|C>(1-<Y|C>)})}.
\end{equation}
The conditions on the default correlations give us the following  recursion 
relations for $p_{n}$ as
\begin{equation} 
p_{n+1}=p_{n}+(1-p_{n})\rho .
\end{equation}
These recursion relations can be solved to give $p_{n}$ as
\begin{equation}
p_{n}=1-(1-p)(1-\rho)^{n}.
\end{equation}
$p_{n}$ increases with $n$ and $p_{n}\to
1$ as $n \to \infty$ for $\rho > 0$. 

From these conditional default probabilities $p_{n} (n=0,\cdots,N-1)$,
the joint default probabilities for the configuration 
$\vec{x}=(x_{1},x_{2},\cdots,x_{N})$
 are given as
\begin{equation}
P(\vec{x})=P(x_{1},x_{2},\cdots,x_{N})
=<\prod_{n=1}^{N}X_{n}^{x_{n}}(1-X_{n})^{1-x_{n}}>  .
\end{equation}
The normalization condition for $P(\vec{x})$ is guaranteed by the 
following decomposition of unity.
\begin{equation}
1=<1>=<\prod_{n=1}^{N} \{ X_{n}+ (1-X_{n}) \} >=
\prod_{j=1}^{N}\big[\sum_{x_{j}=0}^{1}\big]
<\prod_{n=1}^{N}X_{n}^{x_{n}}(1-X_{n})^{1-x_{n}}>
\end{equation}
The probability for $n$ defaults is 
\begin{eqnarray}
P_{N}(n)&=
{}_{N}C_{n}\times P(1,1,\cdots,1,0,\cdots,0)=
{}_{N}C_{n}<\prod_{i=1}^{n}X_{i}
\prod_{i=n+1}^{N}(1-X_{i})>   \nonumber \\
&=
{}_{N}C_{n}
\sum_{k=0}^{N-n}{}_{N-n}C_{k}
(-1)^{k}
(\prod_{n'=0}^{n+k-1}p_{n'}).  \label{P_{N}(n)} 
\end{eqnarray}

We modify the above MCB model as follows. In the MCB model, the 
default correlation is set to be constant irrespective of the number of
default (see\eref{corr_MCB}). We change  the condition as 
\begin{equation} 
\mbox{Cor}(X_{n+1},X_{n+2}|\prod_{n^{'}=1}^{n}X_{n^{'}}=1)=\rho 
\exp(-n \lambda).  \label{decay}
\end{equation}
Here, we introduce a parameter $\lambda >0$ and the default correlation 
 under $n$ defaults decay as $\exp (-n \lambda)$. If we set $\lambda=0$,
 the modified model reduces to the original MCB model.

There are two motivations for the modification.
 The first one is that 
it is mathematically necessary to couple multiple MCB models.
We discuss the mechanism in greater detail 
in the next section. Here, we only comment on the limit value of 
$p_{n}$ as $n \to \infty$.
The modification changes the recursive relation  for $p_{n}$
to 
\begin{equation}
p_{n+1}=p_{n}+(1-p_{n}) \rho \exp (-n \lambda) \label{recursive}. 
\end{equation}
$p_{n}$ is
 calculated as
\begin{equation}
p_{n}=1-(1-p)\prod_{n'=0}^{n-1}(1-\rho_{n'}) \label{solution}  .
\end{equation}
Here $\rho_{n}$ is defined as 
$\rho_{n}=\rho \exp (-n \lambda)$.

\begin{figure}[htbp]
\begin{center}
\includegraphics[width=10.0cm]{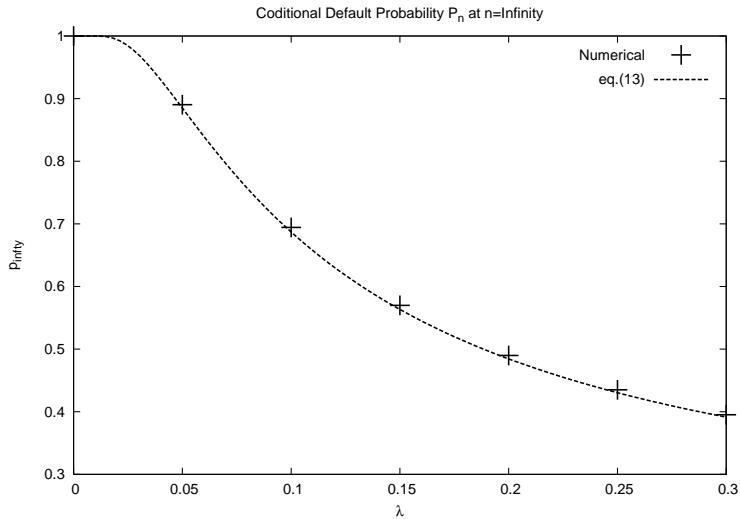}
\caption{$p_{\infty}$ for $0 \le \lambda \le 0.3$ with $p=\rho=0.1$.
The solid line comes from eq.\eref{p_infty_th}. The numerically
enumerated data are shown with + symbol.}
\label{p_infty}
\end{center}
\end{figure}

$p_{n}$ increases with $n$, 
however the increase is reduced by
the decay of the correlation with $n$. The limit
value of $p_{n}$  with   
$n \to \infty$  is roughly estimated as
\begin{equation}
p_{\infty}=\lim_{n\to \infty}p_{n}=1-(1-p)(1-\rho)
\exp (-\rho \frac{e^{-\lambda}}{1-e^{-\lambda}}) . \label{p_infty_th}
\end{equation}
For
$\lambda=0$, $p_{\infty}=1$ and  $p_{\infty}=
p+(1-p)\rho=p_{1}$ for $\lambda=\infty$.
In \Fref{p_infty}, we show the enumerated data for $p_{\infty}$
and the results from eq.\eref{p_infty_th}. As $\lambda$ increases, 
$p_{\infty}$ decreases and the $\lambda$ dependence is well described by 
eq.\eref{p_infty_th}.

\begin{figure}[htbp]
\begin{center}
\includegraphics[width=10.0cm]{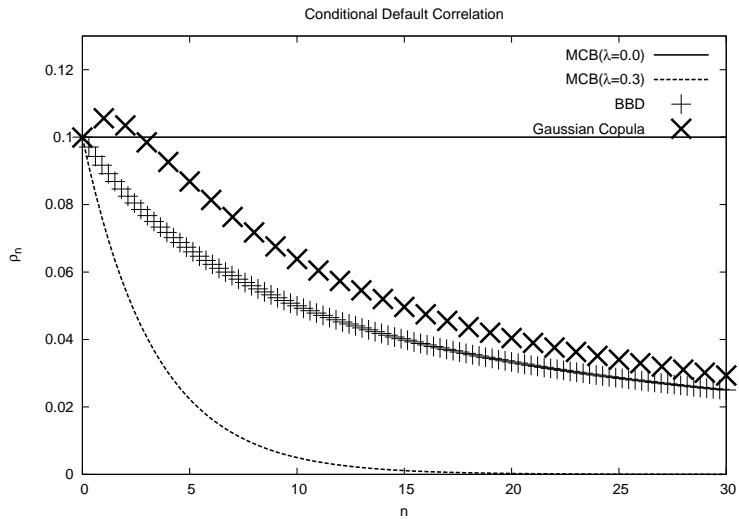}
\caption{$\rho_n$ for MCB (solid, dotted), BBD (+) and Gaussian copula 
models ($\times$).
We set $p=\rho=0.1$ and $N=50$.}
\label{rho_n_compare}
\end{center}
\end{figure}

The second motivation is that 
popular CDO pricing models have decaying correlation with $n$.  
The BBD model's $\rho_n$ is given as \cite{Hisakado}
\begin{equation}
\rho_{n}=\frac{\rho}{1+n\rho}. \label{BBD}
\end{equation}
In the  Gaussian copula model, we do not have 
the explicit form for $\rho_{n}$. From its aggregate loss distribution
function, it is possible to estimate them. 
In \Fref{rho_n_compare}, 
we show $\rho_n$ for the MCB, BBD and Gaussian copula models.
We set $p=\rho=0.1$ and $N=30$.
The Gaussian copula's $\rho_n$ does not show monotonic dependence on
$n$. After a small peak,. it decays to zero. In order to
 mimic the Gaussian copula model within the framework of 
a correlated binomial model, such a dependence should be 
incorporated in the assumption on $\rho_{n}$.

\begin{figure}[htbp]
\begin{center}
\includegraphics[width=10.0cm]{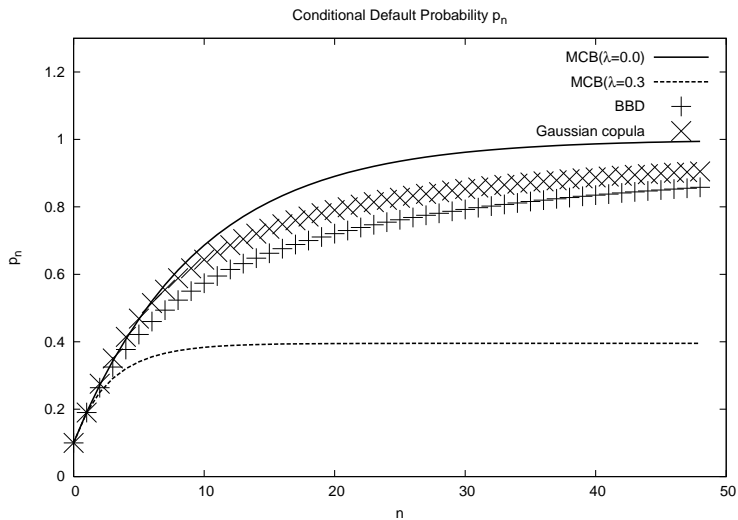}
\caption{$p_n$ for MCB (solid, dotted), BBD (+) and Gaussian copula 
models ($\times$).
We set $p=\rho=0.1$ and $N=50$.}
\label{p_n_compare}
\end{center}
\end{figure}

\Fref{p_n_compare} depicts $p_{n}$ in the same setting.
 With the same $p$  and $\rho$, all $p_{n}$ curves pass through 
$p_{0}=p$ at $n=0$ and $p_{1}=p+(1-p)\rho$ at $n=1$. 
After $n >1$, the behaviors of 
$p_{n}$ depend on the models' definitions on $\rho_{n}$.
$p_{n}$ saturate to about $0.4$ for MCB$(\lambda=0.3)$, which means
that a large scale avalanche does not occur and the loss distribution
function has a short tail.      In MCB with 
$\lambda=0.0$ and
 Gaussian copula models, their $p_{n}$ saturate to 1.
The behaviors are reflected in the fat and long tails in their 
loss distribution.

\begin{figure}[htbp]
\begin{center}
\includegraphics[width=15.0cm,angle=0]{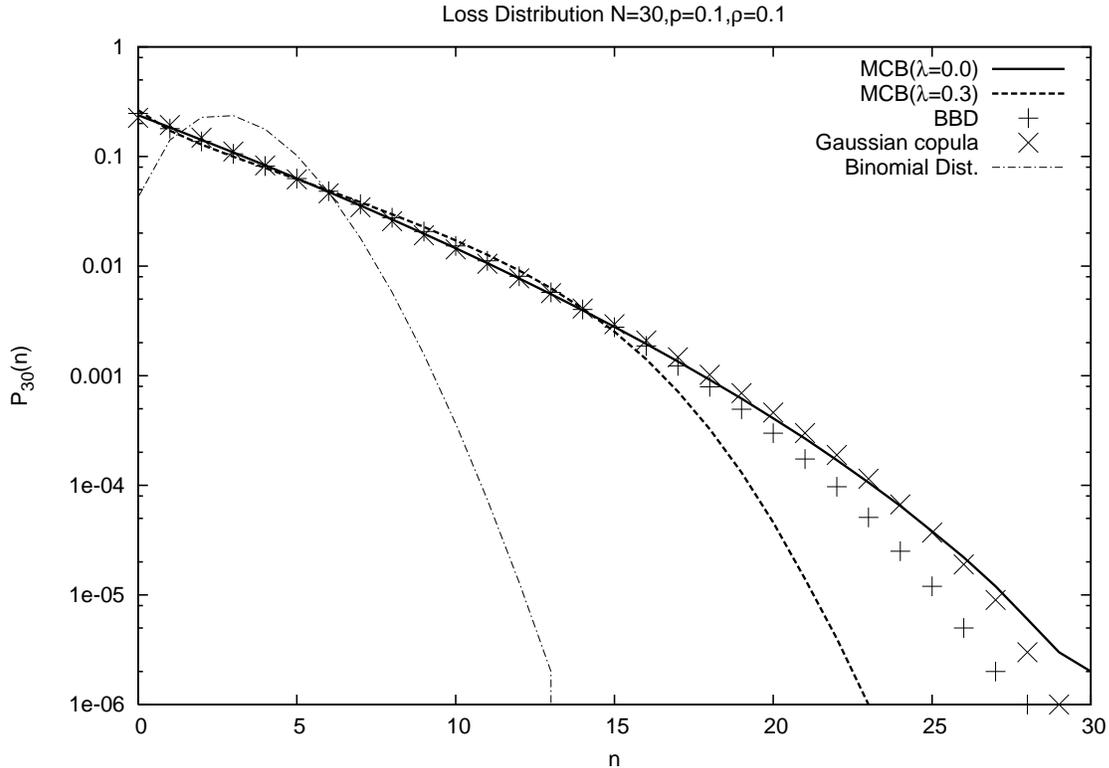}
\caption{Loss Distribution $P_{30}(n)$ for $p=0.1$ and $\rho=0.1$.
$\lambda=0.0,0.3$ and $p=\rho=0.1$.
We also plot the profile for the binomial distribution.}
\label{Dist}
\end{center}
\end{figure}

\Fref{Dist} shows the semi-log plot of 
$P_{30}(n)$ for the MCB, BBD and 
Gaussian copula models.  
We also plot the binomial distribution Bin$(30,0.1)$. The default 
correlation shifts the peak of the binomial distribution 
to $n=0$ and $P_{30}(n)$
comes to have a long tail. 
MCB, BBD and Gaussian copula have almost the same bulk shape.
In particular, in MCB, even if we change $\lambda$, 
$P_{30}(n)$ has almost the same shape for $n  \le 15$. 
 The  bulk  shape of 
$P_{N}(n)$ is mainly determined by $p_{n}$ with small $n$.
$p_{n}$s with large $n$ comes from very rare events 
$\prod_{n^{'}=1}^{n}X_{n^{'}}=1$ and 
contains  information
about  the tails of the distributions.
They do not 
affect the bulk part significantly .

There are differences in their tails.
One sorts the models in the order of thinnest tail to fattest tail,
 we have
\[
\mbox{MCB}(\lambda=0.3) < \mbox{BBD} < \mbox{Gaussian copula} <
\mbox{MCB}(\lambda=0.0)  . 
\]
MCB($\lambda=0$) has almost the same shape as the Gaussian copula.
However, 
it has a  bigger tail than the gaussian coupla at $n=30$. 
The tail of MCB$(\lambda=0.3)$ is short
 compared with other models. 
We can understand this behavior from the behavior of $p_{n}$.

 We also note another role of the damping parameter $\lambda$.
In the calculation $P_{N}(n)$, there are many cancellations 
$\sum_{k=0}^{N-n}{}_{N-n}C_{k}(-1)^{k}$
in  \eref{P_{N}(n)} from the decomposition of
$\prod_{i=n+1}^{N}(1-X_{i})$ . 
This causes numerical errors in the evaluation 
and it is difficult to get $P_{N}(n)$ for $N \ge 40$, even 
if we use long double precision variables in the numerical implementation.
When we set $\lambda=0.3$, the numerical error diminishes greatly and we can 
obtain $P_{N}(n)$ even for $N=100$. This point is 
important when one uses the MCB model for analysis of the actual CDOs 
that have at least 50 assets. 
In addition, with $\lambda>0$, we can take
$\rho$ to be negatively large enough. In S\&P's data, 
a negative default correlation of $0.1\%$ or so has been reported
\cite{SP}.      
We think that
this point is also an  advantage of the modified model.

Hereafter, we mainly focus on the generalization of the MCB model.
However, the same method and reasoning should be applicable to 
other correlated binomial models with any assumption on $\rho_{n}$. 
 If we set $\rho_{n}$ as in \eref{BBD}, we have Beta-Binomial
default distribution models for inhomogeneous portfolios.

\section{Generalization to Inhomogeneous Portfolios}
\label{model}

In this section we couple multiple MCB models and construct 
the joint default probabilities and $P_{N}(n)$ for inhomogeneous portfolios.
In addition, we show that the inter-sector default correlation 
can be set to be large enough by choosing $\lambda$ and other parameters. 
We think that  it is possible to use  the model 
as a model for portfolio credit risk.

\subsection{Coupling of $X$ and $Y$: 1+1 MCB model}
\label{1+1}

Before proceeding to the coupling of multiple MCB models, 
we recall some results for the coupling of 
two random  variables $X$ and $Y$. The default probability is $p_{x}$ 
($p_{y}$) for $X$ (respectively for $Y$) and the default correlation
 between them is $\rho_{xy}$, 
\begin{equation}
\mbox{Corr}(X,Y)=\rho_{xy}.
\end{equation}
As in the MCB model, we introduce the conditional default 
probabilities as
\begin{eqnarray}
&p_{0}=p_{x}=<X> \hspace*{0.3cm}\mbox{and}\hspace*{0.3cm}
p_{1}=<X|Y=1>. \nonumber \\
&q_{0}=p_{y}=<Y> 
\hspace*{0.3cm}\mbox{and}\hspace*{0.3cm} q_{1}=<Y|X=1>.
\end{eqnarray}
From the default correlation $\rho_{xy}$, 
$p_{1}$ and $q_{1}$ are calculated as
\begin{eqnarray}
&p_{1}=p_{0}+(1-p_{0})\sqrt{\frac{p_{0}(1-q_{0})}{(1-p_{0})q_{0}}}
\rho_{xy} \nonumber \\
& q_{1}=q_{0}+(1-q_{0})
\sqrt{\frac{q_{0}(1-p_{0})}{(1-q_{0})p_{0}}}\rho_{xy}.
\end{eqnarray}
In the symmetric (homogeneous) case $p_{x}=p_{y}$, 
the equality $p_{1}=q_{1}$ holds
 and they are given as 
\[
p_{1}=q_{1}=p_{x}+(1-p_{x})\rho_{xy}.
\]
The correlation $\rho_{xy}$ can be set to be $1$ and in the limit 
$p_{1}=q_{1} \to 1$. The maximum value of $\rho_{xy}$ is 1 in the
symmetric case. Conversely in the asymmetric case ($p_{x}\neq
p_{y}$), $\rho_{xy}$ cannot set to be $1$. The maximum value of
$\rho_{xy}$ is determined by the condition that $p_{1} \le 1$ and
$q_{1}\le 1$. From these conditions, we derive the following conditions 
\cite{Lucas}
on $\rho_{xy}$ as
\begin{equation}
\rho_{xy} \le \sqrt{\frac{q_{0}(1-p_{0})}{(1-q_{0})p_{0}}}
\hspace*{0.3cm}\mbox{and}\hspace*{0.3cm}
\rho_{xy}\le \sqrt{\frac{p_{0}(1-q_{0})}{(1-p_{0})q_{0}}}.
\end{equation}
We introduce an asymmetric parameter $r$ as 
\begin{equation}
r=\frac{p_{y}}{p_{x}}
\end{equation}
and a function $f(r,p)$ as
\begin{equation}
f(r,p)=\sqrt{\frac{(1-p)r}{1-rm}}.
\end{equation}
The maximum value of $\rho_{xy}$ is then given as
\begin{equation}
\mbox{Max}(\rho_{xy})=\mbox{Min}(f(r,p_{0}),f(r,p_{0})^{-1}). \label{Max}
\end{equation}
Here Max$(\rho_{xy})$ represents the maximum value of $\rho_{xy}$ and
Min$(A,B)$ means that the smaller value of $A$ and $B$ is taken.
\begin{figure}[htbp]
\begin{center}
\includegraphics[width=10.0cm,angle=0]{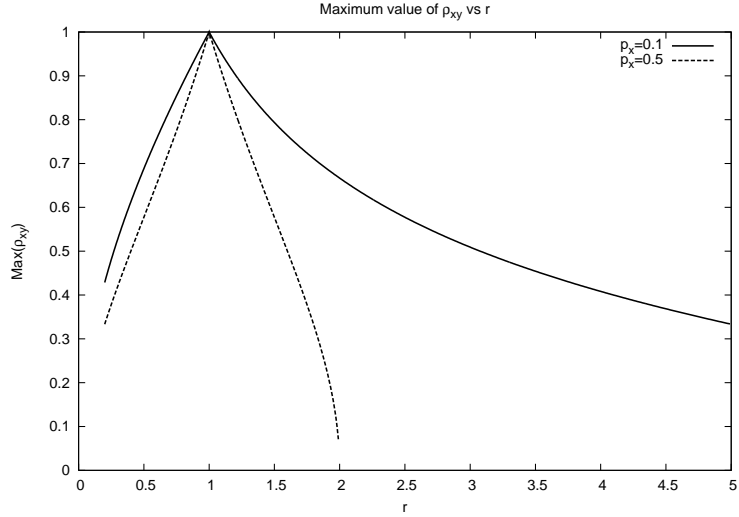}
\caption{
Max$(\rho_{xy})$ for $p_{x}=0.1$(solid line), $0.5$(dotted line) and $0.2 \le r \le 5.0$.
}
\label{Max_rho_xy}
\end{center}
\end{figure}
\Fref{Max_rho_xy} shows Max$(\rho_{xy})$ as a function of the 
asymmetric  (inhomogeneity) parameter $r$. We show two curves, the solid
one for $p_{x}=0.1$ and the dotted one for $p_{x}=0.5$.
As the inhomogeneity $r$ increases, that is $r$ departs from $r=1$,
Max$(\rho_{xy})$ decreases.  
For fixed $r$, as $p_{x}$ becomes large, Max$(\rho_{xy})$
becomes small. The reason is that the condition $p_{1}\leq 1$ becomes 
more difficult to satisfy
as $p_{x}=p_{0}$ increase. $p_{1}$ is a monotonous increasing function
of $p_{0}$.
In the previous section, the conditional default probability 
$p_{n}$ becomes smaller as we set  $\lambda$ larger.
When  we set a large $\lambda$, we show that it is possible to couple 
multiple MCB models with strong default correlation.

\subsection{$N+1$ MCB model}

For the second step, we couple an  $N$ assets 
MCB model with one two-valued random variable $Y$.
We introduce $N$ random variables $X_{n} (n=1,\cdots,N)$ and 
the default probability and the default correlation for them is
$<X_{n}>=p_{x}$ and $\rho$. The default probability for $Y$ is $<Y>=p_{y}$
and the default correlation between $Y$ and $X_{n}$ is written as 
$\rho_{xy}$.
We assume homogeneity for the $N$ assets MCB model  and the default correlation
between $X_{n}$ and $Y$ is independent of the asset index
 $n$.
As in the previous cases, we introduce conditional 
default probabilities as
\begin{eqnarray}
&p_{n,0}=<X_{n+1}|\prod_{n^{'}=1}^{n}X_{n^{'}}=1>  
\hspace*{0.2cm}\mbox{and} \hspace*{0.2cm} 
p_{n,1}=<X_{n+1}|\prod_{n^{'}=1}^{n}X_{n^{'}}\times Y=1>
\nonumber \\
&q_{n}=<Y|\prod_{n^{'}=1}^{n}X_{n^{'}}=1>.
\end{eqnarray}
The joint default probabilities $P(\vec{x},y)=P(\vec{X}=\vec{x},y)$
are calculated by decomposing the following expression with
these conditional
default probabilities  
\begin{equation}
P(\vec{x},y)
=<\prod_{n=1}^{N}X_{n}^{x_{n}}(1-X_{n})^{1-x_{n}}\times Y^{y}(1-Y)^{1-y}> 
\end{equation}
The joint default probabilities with the condition  $Y=1$ are
\begin{equation}
P(\vec{x}|Y=1)
=<\prod_{n=1}^{N}X_{n}^{x_{n}}(1-X_{n})^{1-x_{n}}>/p_{y}.
\end{equation}
The joint probabilities with the condition $Y=0$ are obtained by 
the following  relations.
\begin{equation}
P(\vec{x}|Y=0)(1-p_{y})+P(\vec{x}|Y=1)p_{y}=
P(\vec{x})
=<\prod_{n=1}^{N}X_{n}^{x_{n}}(1-X_{n})^{1-x_{n}}>
\end{equation}
The explicit form for $P(\vec{x}|y=0)$ is
\begin{equation}
P(\vec{x}|Y=0)=\frac{P(\vec{x})-P(\vec{x}|Y=1)p_{y}}{1-p_{y}}.
\end{equation}
The probability for $n$ defaults is 
given as in eq.\eref{P_{N}(n)}
\begin{equation}
P_{N}(n)={}_{N}C_{n}\sum_{k=0}^{N-n}{}_{N-n}C_{k}
(-1)^{k}
(\prod_{n'=0}^{n+k-1}p_{n',0}). 
\end{equation}
The probability for $n$ defaults with $Y=1$ is
\begin{equation}
P_{N}(n|Y=1)={}_{N}C_{n}\sum_{k=0}^{N-n}{}_{N-n}C_{k}
(-1)^{k}
(\prod_{n'=0}^{n+k-1}p_{n',1}). 
\end{equation}
Using the same argument as for the joint probabilities with $Y=0$,
$P_{N}(n|Y=0)$ is written as
\begin{equation}
P_{N}(n|Y=0)=\frac{P_{N}(n)-P_{N}(n|1)p_{y}}{1-p_{y}}.
\end{equation}

About the conditional default probabilities
$p_{n,0}$, we impose the same conditions 
 as with the homogeneous $N$ assets MCB model.
\begin{equation}
\mbox{Cor}(X_{n+1},X_{n+2}|\prod_{n^{'}=1}^{n}X_{n^{'}}=1)=
\rho e^{-n \lambda}
\end{equation}
The same recursive relation \eref{recursive} for $p_{n,0}$ is obtained
 and the $p_{n,0}$ is given by
\begin{equation}
p_{n,0}=1-(1-p_{x})\prod_{n'=0}^{n-1}(1-\rho_{n'}) \label{solution2}  .
\end{equation}
Here $\rho_{n}$ is defined as before.
Fot  the conditions on $p_{n,1}$ and $q_{n}$, there are two possible ways to realize  a strong correlation.
The only way to realize a strong correlation $\rho_{xy}$
between $X_{n+1}$ and $Y$ is
\begin{equation} 
\mbox{Corr}(X_{n+1},Y|\prod_{n^{'}=1}^{n}X_{n^{'}})=
\rho_{xy} e^{-n \lambda} \label{N+1}.
\end{equation}
In this case, the relations for $p_{n,1}$ and $q_{n}$ are
\begin{equation}
p_{n,1}q_{n}=q_{n+1}p_{n,0}=p_{n,0}q_{n}+\rho_{xy}
\sqrt{p_{n,0}(1-p_{n,0})q_{n}(1-q_{n})}
e^{-n \lambda}.
\end{equation}
The recursive relations are
\begin{eqnarray}
p_{n,1}&=p_{n,0}+\rho_{xy}e^{-n \lambda}(1-p_{n,0})
\sqrt{\frac{p_{n,0}(1-q_{n})}{(1-p_{n,0})q_{n}}}
  \\
q_{n+1}&=q_{n}+\rho_{xy}e^{-n \lambda}(1-q_{n})
\sqrt{\frac{(1-p_{n,0})q_{n}}{p_{n,0}(1-q_{n})}}.
\end{eqnarray}
If we set $p_{y}=p_{x}$ and $\rho_{xy}=\rho$, these relations
reduce to $p_{n,1}=p_{n+1,0}$ and $q_{n}=p_{n,0}$ and this coupled  
model is nothing but the $(N+1)$ assets MCB model.

\begin{figure}[h]
\begin{center}
\includegraphics[width=12.0cm,angle=0,clip]{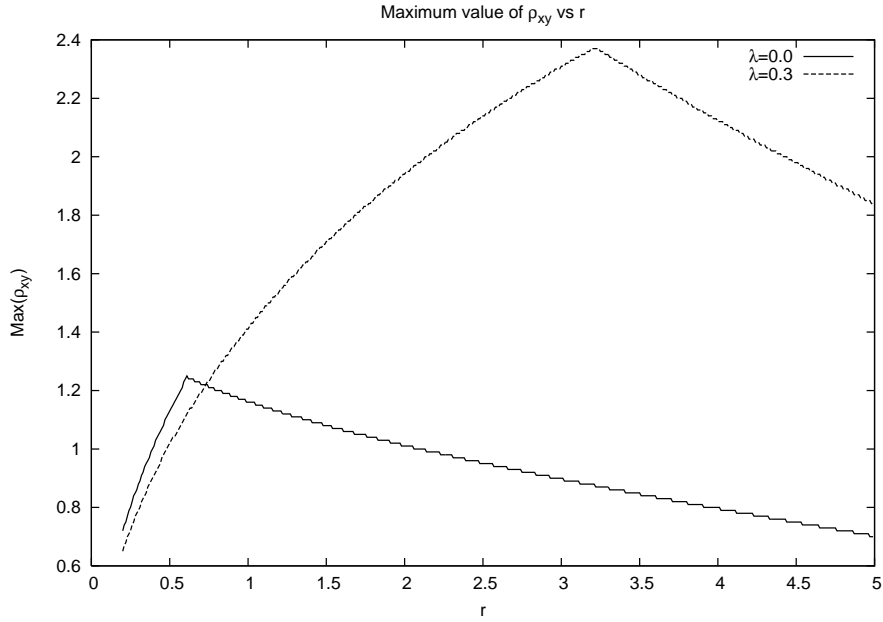}
\caption{
Max($\rho_{xy}$)/$\rho$ vs $r=\frac{p_{y}}{p_{x}}$.
$N=30$, $\rho=0.1$, $p_{x}=0.1$. The solid line $\lambda=0.0$
and the dotted line $\lambda=0.3$}
\label{max_N1}
\end{center}
\end{figure}

We write  the ratio $\frac{q_{n}}{p_{n,0}}$ as $r_{n}$ and  the 
conditions that $p_{n,1}\le 1$ and $q_{n+1}\le 1$ are summarized as
\begin{eqnarray}
\rho_{xy} \le e^{n \lambda} \times 
\mbox{Min}
(f(r_{n},p_{n,0})
,f(r_{n},p_{n,0})^{-1}). \label{Max2}
\end{eqnarray}
Min$(f(r_{n},p_{n,0}),f(r_{n},p_{n,0})^{-1})$ 
is nothing but the condition
for Max$(\rho_{xy})$ of  the two random variables $X,Y$ with 
default probabilities $(p_{x},p_{y})=(p_{n,0},r_{n}p_{n,0})$ (see
eq.\eref{Max}). As explained above,
Min$(f(r_{n},p_{n}),f(r_{n},p_{n})^{-1})$ takes  maximum value $1$
at $r_{n}=1$ for any value of $p_{n}$. 
It also decrease with the increase of $p_{n}$ 
for fixed $r_{n}$.

The necessary condition for the model to be  self-consistent is that 
$p_{n,1}\le 1$ and $q_{n}\le 1$ for all $n$. We discuss $\lambda=0$ 
and $\lambda>0$ cases separately.
\begin{itemize}
\item $\lambda=0.0$: $p_{n,0}$ increases with $n$ 
and $p_{n,0} \to 1$ as $n \to \infty$. 
The range of $p_{n,0}$ is $[p_{x},p_{N-1}] \simeq [p_{x},1]$ 
and it is difficult
to choose $p_{y}$ such that $r_{n} \simeq 1$ for all $n$.
If $r_{n}$ departs much  from 1 for some $n$, 
Max$(\rho_{xy})$ decreases. The choice $p_{x}=p_{y}$ and 
$\rho_{xy}=\rho$ is possible, we anticipate that 
 Max$(\rho_{xy})$ decreases from
$\rho$ as $p_{y}$ departs from $p_{x}$. 

\item $\lambda >0$: The limit value of $p_{n}$ 
becomes small (see eq.\eref{p_infty_th}) and
the range of $p_{n}$ is narrow as compared with the $\lambda=0.0$ case. 
It may be possible to choose $p_{y}$ such that
the asymmetric parameter $r_{n}$ is small for small $n$. 
In addition, for large $n$, it is easy to satisfy the
condition \eref{Max2} because 
of the prefactor $\exp(n \lambda )$. We think
 that  Max$(\rho_{xy})$ is large in this case. 
\end{itemize}

We have checked numerically the values of the joint probabilities for
all configurations $(\vec{x},y)$ with $N=30$ and $p_{x}=\rho=0.1$.
In \fref{max_N1}, we show the data of Max$(\rho_{xy})/\rho$ for the 
case  $\lambda=0.0$ (solid line) and $\lambda=0.3$ (dotted line).
In the $\lambda=0.0$ case,
Max$(\rho_{xy}) \simeq \rho$ near $r=1$ and as $r$ departs from 
1, Max$(\rho_{xy})$ decreases from $\rho$.
The data for $\lambda=0.3$ case show that it is possible 
to set  a large Max$(\rho_{xy})$  if we use a large $r$.  
We can set  $\rho_{xy}$ as strong as several times of $\rho$.
  
We also point out that the above $N+1$ MCB model can be used to
describe a credit portfolio where one obligor has great exposure.
Such an obligor is described  by $Y$ and other obligors are
by $X_{n}$. If $p_{y}$ is quite different from $p_{x}$, we 
can couple $Y$ and $X_{n}$ with strong $\rho_{xy} \simeq \rho_{x}$ 
by setting a sufficiently large  $\lambda$.

\subsection{$N+M$ MCB Model : Coupled MCB model}

\begin{figure}[htbp]
\begin{center}
\includegraphics[width=10.0cm,angle=0,clip]{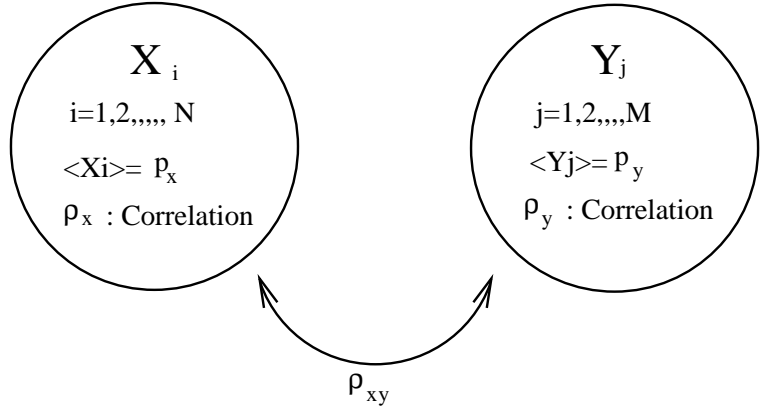}
\caption{
Portfolio of $N+M$ MCB model. The first (second) sector has 
parameters $N (M)$,$p_{x}$ $(p_{y})$ and $\rho_{x} (\rho_{y})$.
The inter-sector default correlation is $\rho_{xy}$
}
\label{two}
\end{center}
\end{figure}

Next we  consider a portfolio with two sectors. The first sector has $N$
assets and the second has $M$ assets. To construct the joint default
probabilities for the portfolio, we try to couple two MCB models. 
The former model's $N$ assets are
described by $X_{n}(n=1,\cdots,N)$ and the states of the latter model's assets
are  described by $Y_{m} (m=1,\cdots,M)$.
The default probability and the default correlation in each sector are
 ($p_{x},\rho_{x}$) and ($p_{y},\rho_{y}$). 
The default correlation between the assets in different sectors 
is denoted as $\rho_{xy}$ (see \Fref{two}).

Introducing the conditional default probabilities 
$p_{n,m}$ and $q_{n,m} $as
\begin{eqnarray}
p_{n,m}&=&<X_{n+1}|\prod_{n'=1}^{n}X_{n'}
\prod_{m'=1}^{m}Y_{m'}=1>,  \hspace*{0.5cm} p_{0,0}=p_{x} \\
q_{n,m}&=&<Y_{m+1}|\prod_{n'=1}^{n}X_{n'}\prod_{m'=1}^{m}Y_{m'}=1>,  
\hspace*{0.5cm} q_{0,0}=p_{y}, 
\end{eqnarray}
we impose the following conditions on $p_{n,0}$ and $q_{0,m}$
\begin{eqnarray}
&&\mbox{Cor}(X_{n+1}X_{n+2}|\prod_{n'=1}^{n}X_{n'}=1)=\rho_{x}
 \exp(-n\lambda_{x})    \label{rhox}  \\
&&\mbox{Cor}(Y_{m+1}Y_{m+2}|\prod_{m'=1}^{m}Y_{m'}=1)=\rho_{y} 
 \exp(-m\lambda_{y}). \label{rhoy} 
\end{eqnarray}
The recursive relations for $p_{n,0}$ and $q_{0,m}$ are
\begin{eqnarray}
&&p_{n+1,0}=p_{n,0}+\rho_{x}
\exp(-n\lambda_{x})  (1-p_{n,0}) \\
&&q_{0,m+1}=q_{0,n}+\rho_{y} 
\exp(-m\lambda_{y})  (1-q_{0,m}).
\end{eqnarray}
Their solutions are, by denoting 
$\rho_{x,n}=\rho_{x} \exp (-n \lambda_{x})$
and $\rho_{y,m}=\rho_{y} \exp (-m \lambda_{y})$,
\begin{eqnarray}
&&p_{n,0}=1-(1-p_{x})\prod_{n'=0}^{n-1}(1-\rho_{x,n'}) \\
&&q_{0,m}=1-(1-p_{y})\prod_{m'=0}^{m-1}(1-\rho_{y,m'}).
\end{eqnarray}
For  the inter-sector correlation, we impose the next conditions
on $p_{n,m}$ and $q_{n,m}$, which is a natural generalization of $N+1$
case (see eq.\eref{N+1}).
\begin{equation}
\mbox{Cor}(X_{n+1}Y_{m+1}|
\prod_{n'=1}^{n}X_{n'}\prod_{m'=1}^{m}Y_{m'}=1)
=\rho_{xy} e^{- (n \lambda_{x}+m\lambda_{y})}.
  \label{corr_2}
\end{equation}
We obtain the following recursive relations,
\begin{eqnarray}
&p_{n,m+1}q_{n,m}=q_{n+1,m}p_{n,m} \nonumber \\
&=p_{n,m}q_{n,m}+\rho_{xy} 
e^{-(n_{x}\lambda_{x}+m\lambda_{y})}
\sqrt{p_{n,m}(1-p_{n,m})q_{n,m}(1-q_{n,m})}. \label{sc_2}
\end{eqnarray}
Using these relations, we are able to calculate  
$p_{n,m}$ and $q_{n,m}$
iteratively starting from $p_{n,0}$ and $q_{0,m}$.

The joint default probability for the portfolio 
configuration $(\vec{x},\vec{y})$
is calculated by decomposing the following expression with $p_{n,m}$
and $q_{n,m}$
\begin{eqnarray}
& P(\vec{x},\vec{y})=P(x_{1},x_{2},\cdots,x_{N},
y_{1},y_{2},\cdots,y_{M}) \nonumber \\
& =<\prod_{n=1}^{N}X_{n}^{x_{n}}(1-X_{n})^{1-x_{n}}
\prod_{m=1}^{M}Y_{m}^{y_{m}}(1-Y_{m})^{1-y_{m}}>. 
\end{eqnarray}
In particular, the probability for $n,m$ defaults in each sector,
 which is denoted as $P_{N,M}(n,m)$, is  
\begin{eqnarray}
&P_{N,M}(n,m)=
{}_{N}C_{m}\times {}_{M}C_{m}\times 
P(
1,1,\cdots,1,0,\cdots,0,
1,1,\cdots,1,0,\cdots,0
) \nonumber \\
&=
{}_{N}C_{m}\times {}_{M}C_{m}\times 
<\prod_{n'=1}^{n}X_{n'}
\prod_{k=n+1}^{N}(1-X_{k})
\prod_{m'=1}^{m}Y_{m'}
\prod_{l=m+1}^{M}(1-Y_{l}) >
\nonumber 
\\
&=
{}_{N}C_{m}\times {}_{M}C_{m}
\times 
\sum_{k=0}^{N-n}\sum_{l=0}^{M-m}
(-1)^{k+l}
{}_{N-n}C_{k} \times
{}_{M-m}C_{l}  \nonumber \\
&\times < \prod_{n'=1}^{n+k}X_{n'}\prod_{m'=1}^{m+l}Y_{m'} >. 
\end{eqnarray}
$P_{N+M}(n)$ is easily calculated from $P_{N,M}(n,m)$ as
\begin{equation}
P_{N+M}(n)=\sum_{n'=0}^{n}P_{N,M}(n',n-n').
\end{equation}
In decomposing $< \prod_{n=1}^{k}X_{n}\prod_{m=1}^{l}Y_{m} >$
, one can do it in any order. 
The independence of the order of the decomposition of
$< \prod_{n=1}^{k}X_{n}\prod_{m=1}^{l}Y_{m} >$
is guaranteed by \eref{corr_2} and \eref{sc_2}.
We decompose it as  
\begin{equation}
< \prod_{n=1}^{k}X_{n}\prod_{m=1}^{l}Y_{m} >
=\prod_{n=0}^{k-1}p_{n,0}\times \prod_{m=0}^{l-1}q_{k,m}. 
\end{equation}

For  the maximum value of $\rho_{xy}$, it is necessary to
check all values of the joint probabilities. However, $N+M$ 
model is reduced to $N+1$ or $1+M$ model by choosing $M=1$ or $N=1$ 
respectively. From  the discussions and the results 
in the previous subsection for
the $N+1$ model, we can anticipate as follows.
\begin{enumerate} 
\item 
$\lambda_{x}=\lambda_{y}=0$: If $p_{x}=p_{y}$ and $\rho_{x}=\rho_{y}$, 
two MCB models are the same  and we can set $\rho_{xy}=\rho_{x}=\rho_{y}$.
Two models merge completely and we have a $(N+M)$ assets 
MCB model. 
As the asymmetry between the two models
becomes large ($p_{x}\neq p_{y}$ or $\rho_{x}\neq \rho_{y}$)
, Max$(\rho_{xy})$ decreases from $\rho_{x},\rho_{y}$.
 
\item 
$\lambda_{x},\lambda_{y}>0$: As
$\lambda_{x},\lambda_{y}$  increase, Max$(\rho_{xy})$
becomes large. The asymmetry in $p_{x},p_{y}$ and $\rho_{x},\rho_{y}$
diminishes Max$(\rho_{xy})$.
\end{enumerate}

\begin{figure}[htbp]
\begin{center}
\includegraphics[width=15.0cm,angle=0,clip]{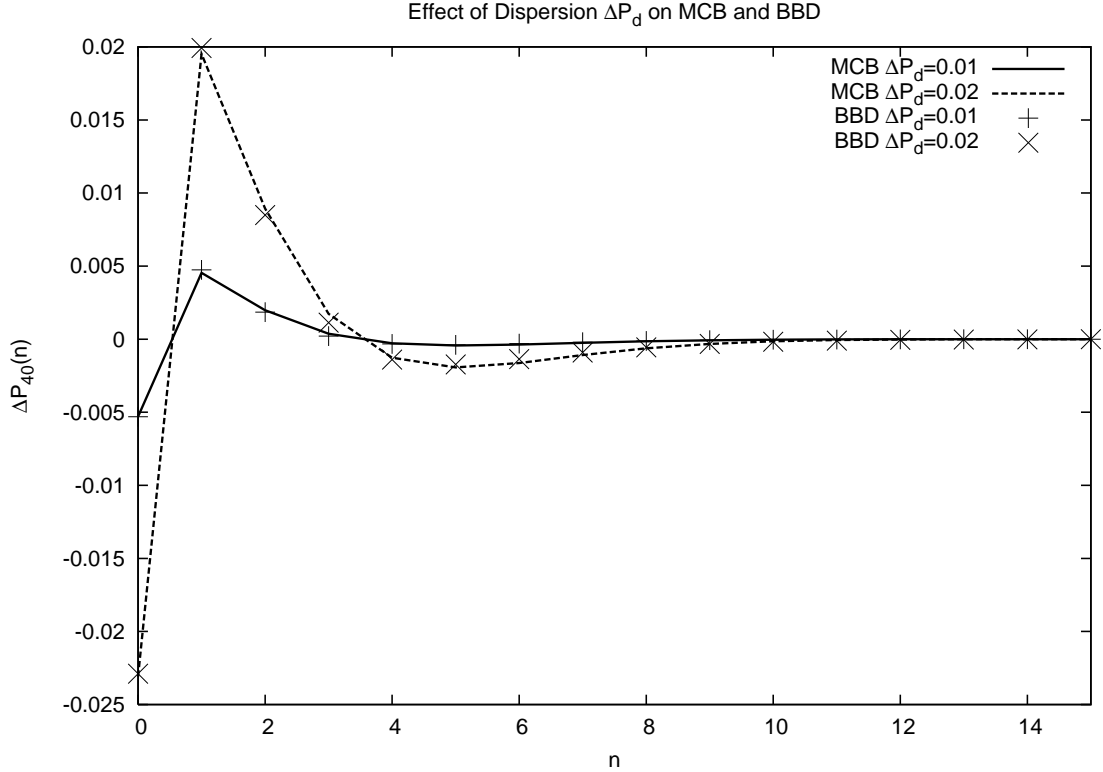}
\caption{$\Delta P(n)$ vs $n$. We set 
$N=20$, $p=0.03$, $\rho=0.03$ and $\lambda_{x}=\lambda_{y}=0.3$.
MCB in solid and dotted lines, and BBD in + and $\times$ symbols.}
\label{disorder}
\end{center}
\end{figure}

Using the above coupled  $N+M$ MCB model, 
we study the effect of the dispersion 
of the default probability on $P_{N}(n)$
and on the evaluation of tranches. 
 More complete analysis about the
difference between the usage of individual spreads and 
of portfolio average spreads in CDO pricing 
has been performed in \cite{Finger2}. 
There, the usage of the average spread results in the lower estimation
of the equity tranche. 
We consider a portfolio
with $N+N$ assets. The assets in each sector have default probabilities
$p \pm \Delta P_{d}$  and an  intra-sector default correlation $\rho$. 
We set the inter-sector default correlation $\rho_{xy}$ also as $\rho$. 
The inhomogeneity in the default probability is controlled by $\Delta
P_{d}$. If we set $\Delta P_{d}=0$, the two sector are completely
merged to one sector and we have a homogeneous $2N$ MCB  model with 
$p$ and $\rho$.  
In \fref{disorder}, we shows the default probability difference $\Delta
P(n)$ between the inhomogeneous case $P_{N+N}(n)$ with $\Delta P_{d}
\neq 0$ and homogeneous case $P_{2N}(n)$.
$\Delta P(n)$ is defined as
\begin{equation}
\Delta P(n)=P_{N+N}(n)-P_{2N}(n).
\end{equation}
We set $N=20$, $p=\rho=0.03$ and $\lambda=0.3$. 
The solid curve represents the data 
for $\Delta P_{d}=0.01$ and the dotted curve stands for the case
$\Delta P_{d}=0.02$.
We see that $\Delta P(n)$ is large only for small $n$. 
 We also plot the results for the $N+M$ BBD
model. To construct the loss distribution function, it is necessary to
change \eref{rhox} and \eref{rhoy} to
\begin{eqnarray}
&&\mbox{Cor}(X_{n+1}X_{n+2}|\prod_{n'=1}^{n}X_{n'}=1)=\rho /(1+n\rho) \\
&&\mbox{Cor}(Y_{m+1}Y_{m+2}|\prod_{m'=1}^{m}Y_{m'}=1)=\rho /(1+m\rho ) .
\end{eqnarray}
Eq.(\ref{corr_2}) is also changed as
\begin{equation}
\mbox{Cor}(X_{n+1}Y_{m+1}|
\prod_{n'=1}^{n}X_{n'}\prod_{m'=1}^{m}Y_{m'}=1)
=\rho /(1+(n+m)\rho).
\end{equation}
The recursive relations are also changed, but the remaining procedures
are the same as for the  $N+M$ MCB model.
As we have shown in the previous section, the bulk shapes of the loss
distribution of MCB and BBD models  are almost the same, and the effects of 
$\Delta P_{d}$ on them are also similar.

\begin{figure}[htbp]
\begin{center}
\includegraphics[width=15.0cm,angle=0,clip]{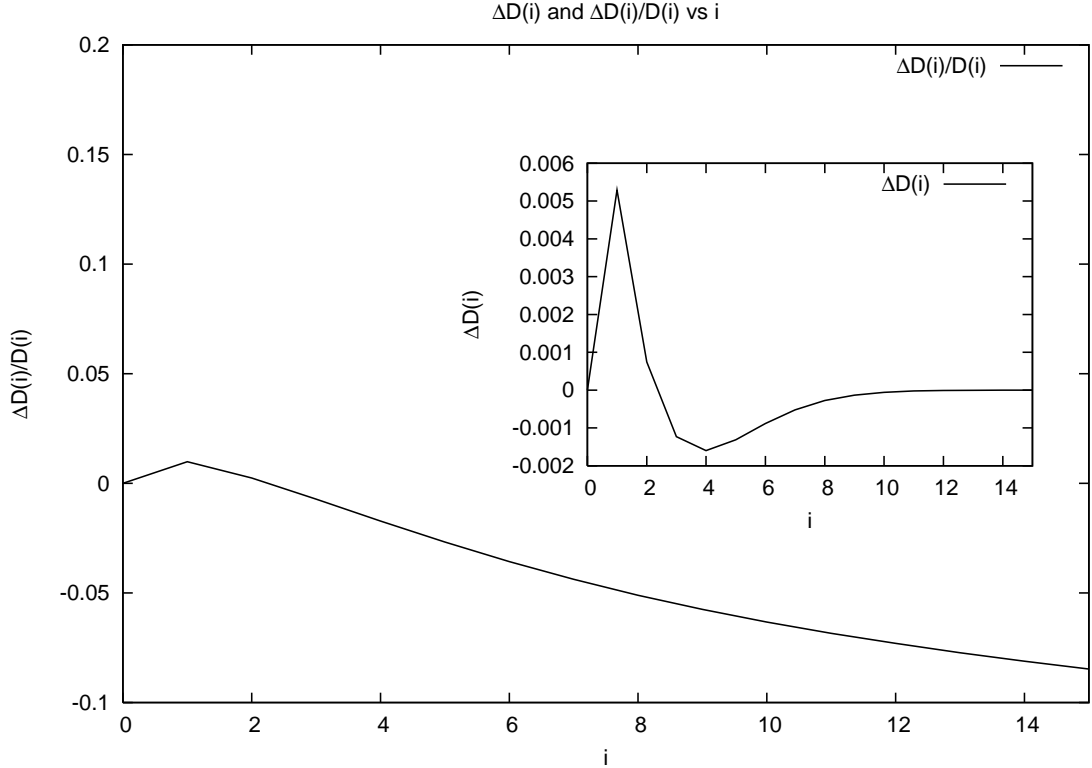}
\caption{Plots of $\Delta D(i)/D(i)$ vs $\Delta P_{d}$ and 
$\Delta D(i)$ vs $\Delta P_{d}$. $N=20$,$p=\rho=0.03$ and $\Delta P_{d}=0.01$.
We plot for $1\le n \le 15$.}
\label{disorder2}
\end{center}
\end{figure}

To see the effect of $\Delta P_{d}$ on the evaluations of  tranches, 
we need to study the
change in the cumulative distribution functions $D(i)$. $D(i)$ is 
defined as
\begin{equation}
D(i)=\sum_{n=i}^{N}P_{N}(n).
\end{equation}
$D(i)$ represents the expected loss rate of the $i-$th tranche, 
which is damaged if more than $i$ assets  default\cite{Mori}. 
One of the important properties of $D(i)$  is
\begin{equation}
\frac{1}{N}\sum_{i=1}^{N}D(i)=p.
\end{equation}
This identity means that the tranches distribute  the portfolio credit risk 
between them. Another important property of $D(i)$ is that the
expected loss rate $D[i,j]$ of the layer protection with attachment
point $i$ and  detachment point $j$ can be built up from $D(i)$ as
\begin{equation}
D[i,j]=\frac{1}{j-i+1}\sum_{n=i}^{j}D(i).
\end{equation}

\Fref{disorder2} shows the plots of $\Delta D(i)$ and of the ratio 
$\Delta D(i)/D(i)$ vs $i$. 
We see that the lower tranches'$(i\le 2)$ expected losses increase by $\Delta
P_{d}$, which is reasonable. The default probability of  
half the assets of the portfolio increases $p\to p+\Delta P_{d}$ 
and the expected losses of the subordinated tranches increase.  
For the senior tranches, the absolute value of $\Delta D(i)$
 decreases with $i$, however this does not mean that $\Delta P_{d}$
does not have a  small effect on them.
The absolute value of $D(i)$ also decreases with $i$.
From \Fref{disorder2}, we see that the ratio $\Delta D(i)/D(i)$
does not necessarily
decrease with $i$.

\subsection{Multi Sector Case}

\begin{figure}[h]
\begin{center}
\includegraphics[width=10cm,angle=0,clip]{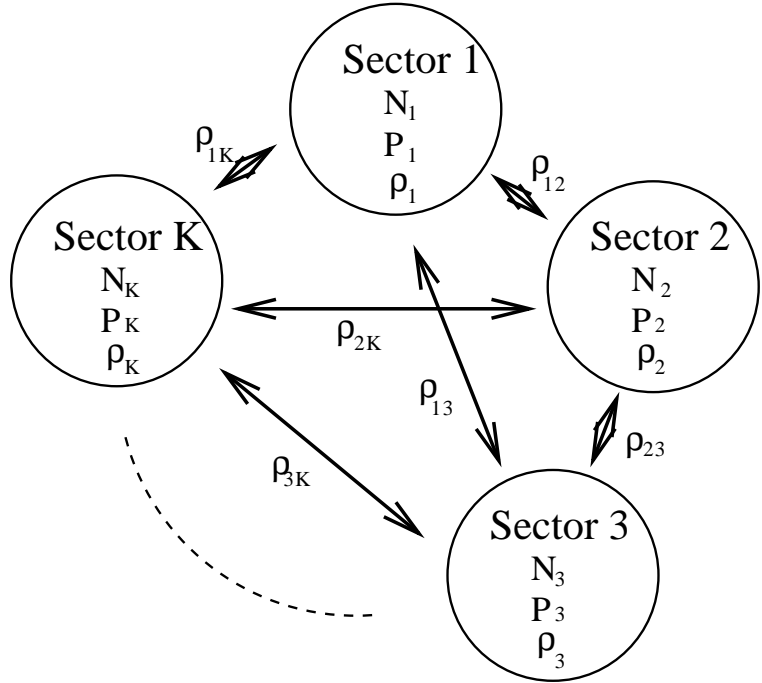}
\caption{Structure of the portfolio.
There are $K$ sectors and $k$-th sector contains $N_{k}$
assets. The default probability and default correlation for
the assets in the $k$-th  sector is $p_{k},\rho_{k}$.
The inter-sector default correlation between the $i-$th 
and $j$-th sector is $\rho_{ij}$.}
\label{multi}
\end{center}
\end{figure}

To couple two or more MCB models,
we consider a portfolio with $N$ assets, which are categorized in 
$K$ different sectors. \Fref{multi} sketches the structure of the portfolio.
 The $k$-th sector contains
$N_{k}$ assets and the relation $\sum_{k=1}^{K}N_{k}=N$ holds.
The states of the assets in the $k$-th sector is described by
$X^{k}_{n_{k}} (n_{k}=1,\cdots,N_{k})$ and 
the default rate and default correlation are denoted as 
$p_{k}$ and $\rho_{k}$.
For the inter-sector default correlation, we denote this as
$\rho_{ij}$ for the default correlation between the $i-$th and $j-$th 
sector. The intra-sector default 
correlation and inter-sector default correlation are 
different and the former is larger than the latter in general
\cite{Jobst}.

We have not yet succeeded in 
the coupling of three or more  MCB models  by generalizing the 
result for the coupled  $N+M$ MCB model.
The reason is that the self-consistency relations are very rigid 
restrictions on the MCB model. These relations, for which the  two-sectors version
is given by \eref{sc_2}, assure the
independence of the order of the decomposition in the estimation
of the expected values of the product of random variables.
It is difficult to  impose  any simple relations on the conditional 
default probabilities that satisfy the self-consistency relations.

\begin{figure}[htbp]
\begin{center}
\includegraphics[width=12cm]{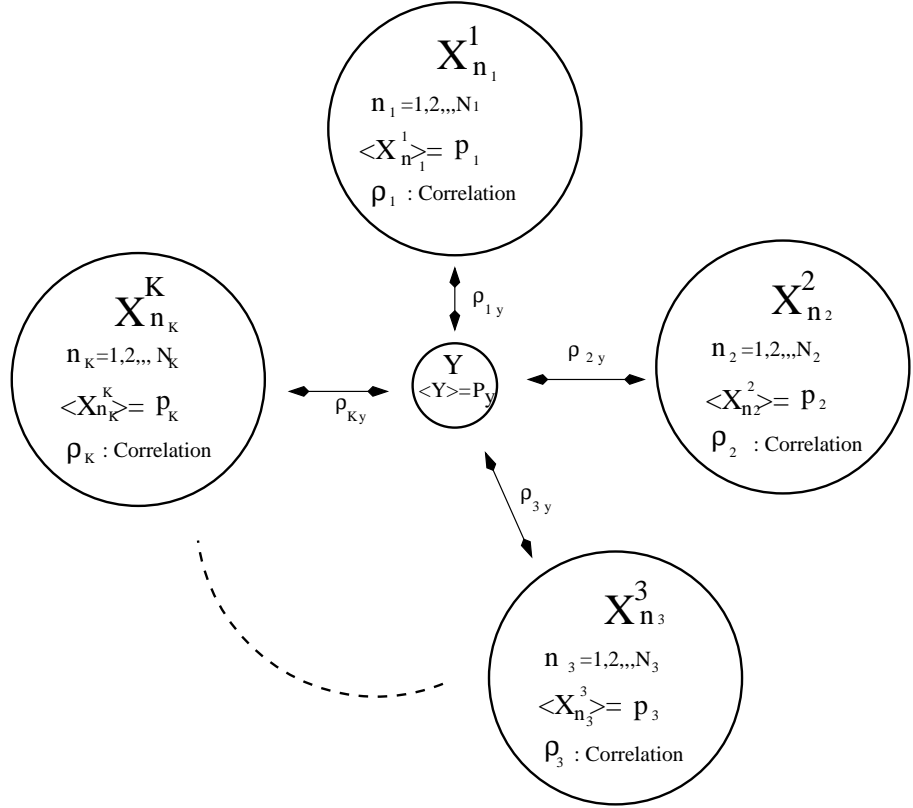}
\caption{Gluing multiple MCB models with $Y$.
The correlation between $X^{k}_{n_{k}}$ and $Y$ is $\rho_{ky}$.
The default correlation between $X^{i}_{n_{i}}$ and $X^{j}_{n_{j}}$
is given as $\rho_{ij}=\rho_{iy}\times \rho_{jy}$.
}
\label{multi_2}
\end{center}
\end{figure}

In order to construct the joint default probabilities
for the assets states $(\vec{x}^{1},\vec{x}^{2},\cdots,\vec{x}^{K})$,
 we do not glue together $K$ MCB models directly.
Instead, as depicted in \fref{multi_2}, we glue multiple MCB models
 through one random variable $Y$. More concretely, we prepare
$K$ sets of $N_{k}+1$ MCB models. 
$N_{k}+1$ MCB model is the  $N_{k}$ MCB model
 coupled with $Y$. The probability of $P(Y=1)$ is written as $p_{y}$.
We Introduce the following conditional default probabilities
\begin{eqnarray}
&p^{k}_{n_{k},0}=<X^{k}_{n_{k}+1}|
\prod_{n_{k}^{'}=1}^{n_{k}}X^{k}_{n_{k}^{'}}=1>  
\hspace*{0.2cm}\mbox{and} \hspace*{0.2cm} 
p^{k}_{n_{k},1}=<X^{k}_{n_{k}+1}|
\prod_{n_{k}^{'}=1}^{n_{k}}X^{k}_{n_{k}^{'}}\times Y=1>
\nonumber \\
&q^{k}_{n_{k}}=<Y|\prod_{n_{k}^{'}=1}^{n_{k}}X^{k}_{n_{k}^{'}}=1>.
\end{eqnarray}
We also  impose the following conditions on 
$p^{k}_{n_{k},0},p^{k}_{n_{k},1}$ and 
$q^{k}_{n_{k}}$ as
\begin{eqnarray}
\mbox{Cor}(X^{k}_{n_{k}+1},X^{k}_{n_{k}+2}
|\prod_{n_{k}^{'}=1}^{n_{k}}X^{k}_{n_{k}^{'}}=1)=
\rho_{k}\exp(-n_{k} \lambda) \\
\mbox{Corr}(X^{k}_{n_{k}+1},Y|
\prod_{n_{k}^{'}=1}^{n_{k}}X^{k}_{n_{k}^{'}})=
\rho_{ky}\exp(-n_{k} \lambda).
\end{eqnarray}
The joint  default probabilities $P(\vec{x}^{k},y)$  and the conditional 
joint default probabilities $P^{k}(\vec{x}^{k}|y)$ are constructed 
as before.
\begin{eqnarray}
&&P(\vec{x}^{k},y)
=<\prod_{n_{k}=1}^{N_{k}} (X^{k}_{n_{k}})^{x^{k}_{n_{k}}}
(1-X^{k}_{n_{k}})^{1-x^{k}_{n_{k}}}\times Y^{y}(1-Y)^{1-y}> 
\\
&&P(\vec{x}^{k},y)=P^{k}(\vec{x}^{k}|y)P(y).
\end{eqnarray}
Packing these conditional default probabilities $P^{k}(\vec{x}^{k}|y)$
 into a bundle, we construct the joint default probabilities for the
 total  portfolio as 
\begin{equation}
P(\vec{x}^{1},\vec{x}^{2},\cdots,\vec{x}^{K})=
\sum_{y=0,1} p(y)\times 
\prod_{k=1}^{K}P(x^{k}_{1},x^{k}_{2},\cdots,x^{k}_{N_{k}}|y).
\end{equation}
We also obtain the default probability function
$P_{N}(n_{1},n_{2},\cdots,n_{K})$ for 
$n_{k}$ default in the $k$-th sector as
\begin{equation}
P_{N}(n_{1},n_{2},\cdots,n_{K})=
\sum_{y=0,1} p(y)\times \prod_{k=1}^{K}P_{N_{k}}(n_{k}|y)
\end{equation}
From the expression, it it easy to calculate the probability 
for $n$ defaults and we write it as $P_{N}(n)$.

For the default correlation between the different sectors, we can show the
next relations.
\begin{equation}
\rho_{ij}=\rho_{iy}\times \rho_{jy}. \label{inter-sector} 
\end{equation}
More generally, the conditional inter-sector default correlations obey
the following relations.
\begin{equation}
\mbox{Cor}(X^{i}_{n_{i}+1},X^{j}_{n_{j}+1}|
\prod_{n_{i}'=1}^{n_{i}}X^{i}_{n_{i}'}
\prod_{n_{j}'=1}^{n_{j}}X^{j}_{n_{j}'}=1
)=\rho_{ij}\exp(-(n_{i}+n_{j}) 
\lambda)  \label{inter-sector2}
\end{equation}
These relations mean that our construction procedure is natural
from the viewpoint of the  original MCB model. In particular, in the  $K=2$ case,
these relations are completely equivalent with those of the $N+M$ coupled
MCB model. See \eref{corr_2} and 
\eref{inter-sector2}. The conditional default probabilities obey the
same conditions.

Furthermore, from the results on Max($\rho_{xy}$) of the  $N+1$ MCB model,
 we see that the model can induce a realistic magnitude of
the inter-sector default correlation.
By choosing $\lambda$ and $p_{y}$ properly, it is possible to set 
$\rho_{ky}$ as large as  several times of $\rho_{k}$.

\begin{figure}[h]
\begin{center}
\includegraphics[width=12.0cm,angle=0,clip]{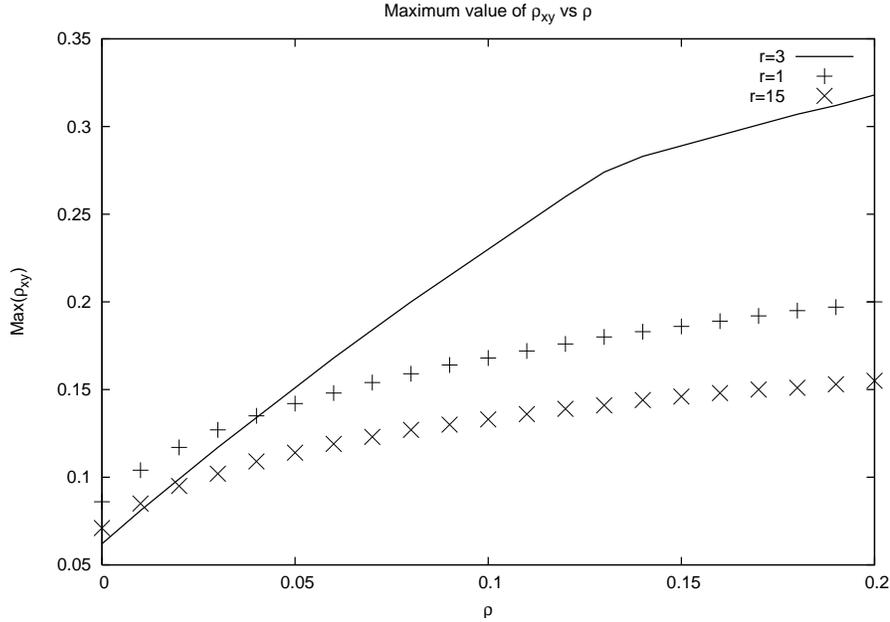}
\caption{
Plot of Max($\rho_{xy}$) vs $\rho$.
$N=30$,$p_{x}=0.03$ and $\lambda=0.3$. }
\label{max_rho_xy_vs_rho}
\end{center}
\end{figure}

\Fref{max_rho_xy_vs_rho} plots Max$(\rho_{xy})$=Max($\rho_{ky}$) 
as functions of $\rho=\rho_{k}$. We set parameters as depicted in the
figure. The solid line depicts the data for 
$r=\frac{p_{y}}{p_{x}}=3.0$. The other two curves correspond to 
$r=1.0$ ($+$) and $r=15.0$ ($\times$).  We see that by setting $r=3.0$.
it is possible to set $\rho_{ky}$ as large as several times of
$\rho_{k}$. If the intra-sector correlation $\rho_{i}=\rho_{j}$ is $10\%$,
we can set $\rho_{iy}=\rho_{jy}=20 \%$. The inter-sector correlations is
then $\rho_{ij}=\rho_{iy}\times \rho_{jy}=4 \%$. In general,
$\rho_{inter}$ is smaller that $\rho_{intra}$, we think that
 the present model can incorporate a strong enough inter-sector default 
correlation.

To prove the relations 
\eref{inter-sector} and \eref{inter-sector2},
in order to calculate the correlation, we need to estimate the 
next expression.
\begin{equation}
<X^{i}_{n_{i}+1}X^{j}_{n_{j}+1}|\prod_{i'=1}^{n_{i}}X^{i}_{i'}\prod_{j'=1}^{n_{j}}X^{j}_{j'}=1>.
\end{equation}
If we fix the random variable $Y$, $X^{i}_{n_{i}}$ and $X^{j}_{n_{j}}$
 are independent. They are coupled by $Y$ and the above equation
is estimated by the average over  $Y=0$ and $Y=1$ as
\begin{eqnarray}
&&
<X^{i}_{n_{i}+1}X^{j}_{n_{j}+1}|
\prod_{i'=1}^{n_{i}}X^{i}_{i'}\prod_{j'=1}^{n_{j}}X^{j}_{j'}=1> 
\nonumber \\
&&=
<X^{i}_{n_{i}+1}|\prod_{i'=1}^{n_{i}}X^{i}_{i'}=1,Y=1> 
<X^{j}_{n_{j}+1}|\prod_{j'=1}^{n_{j}}X^{j}_{j'}=1,Y=1>p_{y}
\nonumber \\
&&+
<X^{i}_{n_{i}+1}|\prod_{i'=1}^{n_{i}}X^{i}_{i'}=1,Y=0> 
<X^{j}_{n_{j}+1}|\prod_{j'=1}^{n_{j}}X^{j}_{j'}=1,Y=0>(1-p_{y})
\nonumber \\
&&=p^{i}_{n_{i},1}p^{j}_{n_{j},1}
p_{y}+\tilde{p}^{i}_{n_{i},0}\tilde{p}^{j}_{n_{j},0}(1-p_{y}). \label{corr_p}
\end{eqnarray}
Here, we denote the conditional default probabilities with the
condition  $Y=0$
as $\tilde{p}^{i}_{n_{i},0}$,
\begin{equation}
\tilde{p}^{i}_{n_{i},0}=<X^{i}_{n_{i}+1}|
\prod_{n_{i}^{'}=1}^{n_{i}}X^{i}_{n_{i}^{'}}\times (1-Y)=1>.
\end{equation}
Between $p^{i}_{n_{i},0}$ and $\tilde{p}^{i}_{n_{i},0}$, the next
relation holds.
\begin{equation}
p^{i}_{n_{i},0}
=p^{i}_{n_{i},1}p_{y}+\tilde{p}^{i}_{n_{i},0}(1-p_{y}).
\end{equation}
In addition, from the correlation between $X^{i}_{n_{i}}$ and $Y$,
we also have the next relations.
\begin{equation}
p^{i}_{n_{i},1}p_{y}=p_{i,0}p_{y}+\rho_{xy}
\sqrt{p_{i,0}(1-p_{i,0})p_{y}(1-p_{y})} 
\end{equation}
Putting these relations into \eref{corr_p}, we can prove the next equations.
\begin{eqnarray}
&<X^{i}_{n_{i}+1}X^{j}_{n_{j}+1}
|\prod_{i'=1}^{n_{i}}X^{i}_{i'}\prod_{j'=1}^{n_{j}}X^{j}_{j'}=1> 
\nonumber \\
&=p^{i}_{n_{i},0}p^{j}_{n_{j},0}
+\rho_{iy}\times \rho_{jy} \sqrt{p^{i}_{n_{i},0}(1-p^{i}_{n_{i},0})
p^{j}_{n_{j},0}(1-p^{j}_{n_{j},0})} \times  e^{-\lambda (n_{i}+n_{j})}
\end{eqnarray}
Using these relations, we calculate the 
inter-sector correlation and prove (\ref{inter-sector2}).

\begin{figure}[htbp]
\begin{center}
\includegraphics[width=15cm]{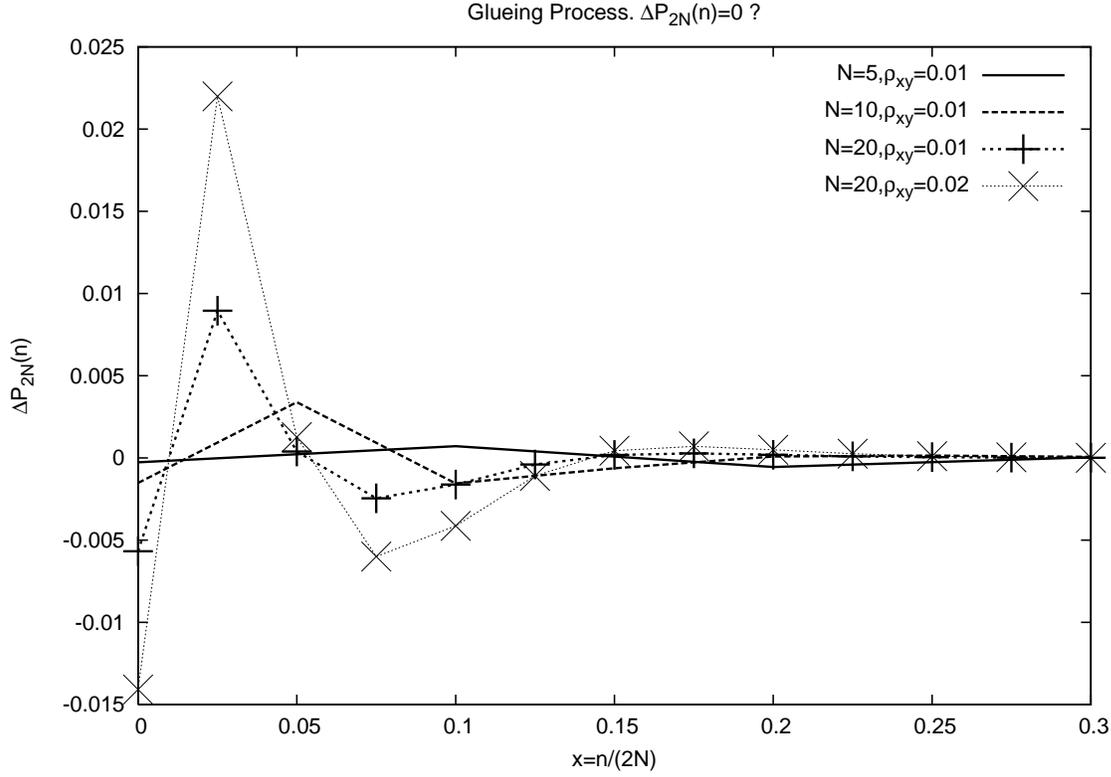}
\caption{Comparison of Two sector MCB model: Directly Coupled
Model vs Coupled by Y Model $(p_{y}=0.5)$.   
Plot of $\Delta P(n)=P_{2N}(n)-P_{N+N}(n)$ vs $n/(2N))$.
$N=5,10,20$,$p=\rho=0.03$ and $\lambda=0.3$. 
Upper three curves have $\rho_{xy}=0.01$
and the bottom line with $\times$ symbol has $\rho_{xy}=0.02$. }
\label{match}
\end{center}
\end{figure}

Next we need to check the validity of the above gluing process.
We consider the two sector case $K=2$ and their intra-sector parameters 
are set to be the same as $N$, $p$ and $\rho$ in each sector. 
For the inter-sector default
correlation $\rho_{12}$, we set $\rho_{1y}^{2}=\rho_{2y}^{2}=\rho_{12}$
in the above multi-sector model.  
If the glueing process of the  multi-sector model works well, 
$P_{2N}(n)$  should coincide with $P_{N+N}(n)$ of the coupled 
$N+N$ MCB model in the 
previous subsection. \Fref{match}  shows 
$\Delta P(n)=P_{2N}(n)-P_{N+N}(n)$ with $N=5,10,20$ and $\lambda=0.3$.
We set $p=\rho=0.03$ and $\rho_{12}=0.01$.
In addition we also plot the data for $\rho_{12}=0.02$ and $N=20$.
 As the system size $N$ becomes large, the discrepancy $\Delta P(n)$
 increases. With the same system size $N=20$, as the inter-sector
correlation increases, the discrepancy also increases.
As we have stated previously, these models obey the same 
conditions on the conditional default probabilities, however  
the default probability profile does not coincide. The glueing process by
the auxiliary random variable $Y$ may cause changes to the joint
probabilities. We have not yet fully understood this point.

With the present model, we study the effects of the inhomogeneous
default correlation $\rho_{inter}\neq \rho_{intra}$ 
on the default distribution function 
$P_{N}(n)$ and the loss rates $D(i)$. We consider ideal portfolios 
which have the same default probability $p_{k}=p$ and default correlation
 $\rho_{k}=\rho$. The inter-sector default correlations are also set to
 be the same value as  $\rho_{ij}=\rho_{inter}$. $N_{k}$ are also
set to be the same $N_{k}=N_s$. We fix the total number of assets as $N
$ and compare $P_{N}(n)$ and $D(i)$ between the portfolios with
different number of sectors $K$. Of course, between $K$ and $N_{s}$ the relation 
$N=N_{s}\times K$ holds. We set $\rho_{inter} < \rho$. 
As the number $K$ increases, the average default correlation
is governed by the inter-sector correlation $\rho_{inter}$ and becomes weak. 

At the extreme limit $K=N$ case, each sector contains
only one asset. In  the inter-sector default correlation, 
$P_{N}(n)$ is given by the superposition of $P_{N}(n|Y=1)$ and 
$P_{N}(n|Y=0)$. The conditional default probabilities $p^{k}_{1,0}$
$p^{k}_{1,1}$ and $\tilde{p}^{k}_{1,0}$ are given as
\begin{eqnarray}
&p^{k}_{1,0}=p \hspace*{0.3cm} \mbox{and}\hspace*{0.3cm}
p^{k}_{1,1}=p+
\sqrt{\rho_{inter}}(1-p)
\sqrt{\frac{p}{1-p}\frac{(1-p_{y})}{p_{y}}} \\
&\tilde{p}^{k}_{1,0}=p-
\sqrt{\rho_{inter}}\hspace*{0.1cm} p
 \sqrt{\frac{1-p}{p}\frac{p_{y}}{(1-p_{y})}} .
\end{eqnarray}
$P_{N}(n|Y=1)$ and $P_{N}(n|Y=0)$ are the binomial
distributions Bin$(N,p^{k}_{1,1})$ and Bin$(N,\tilde{p}^{k}_{1,0})$.

\begin{figure}[htbp]
\begin{center}
\includegraphics[width=14cm]{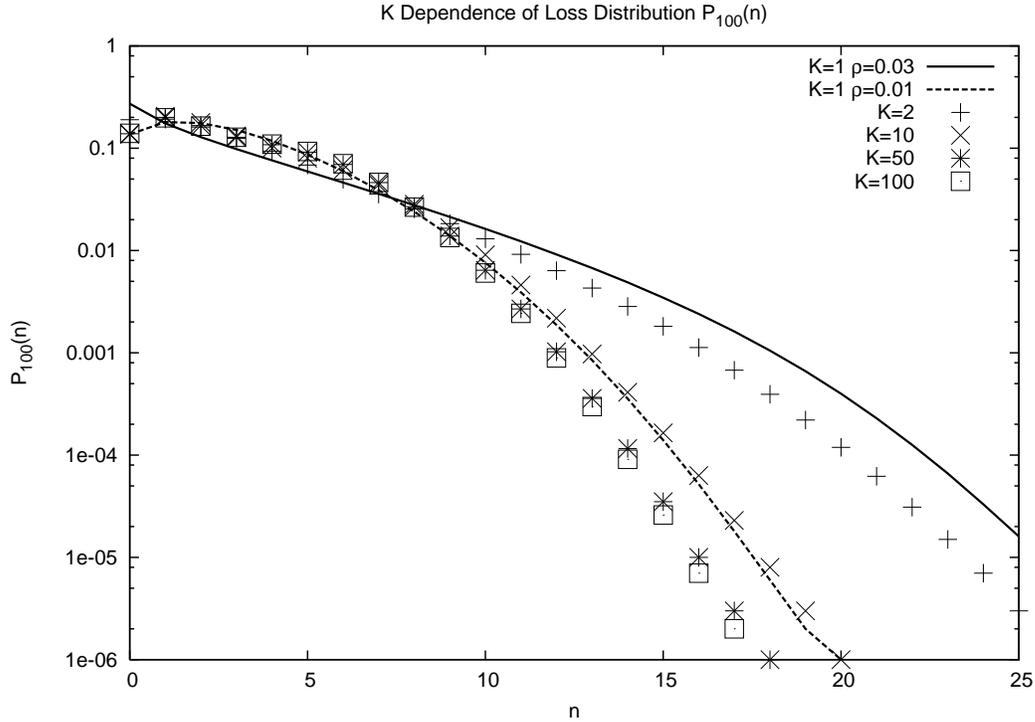}
\caption{Semi-log plot of $P_{N}(n)$ vs $n$.
$N=100$, $K=2,10,100$, $p=\rho=0.03$, $\rho_{inter}=0.01$ 
and $\lambda=0.3$. We set $p_{y}=0.5$.
Solid line represents $K=1$ and $\rho=0.03$ case and the dotted line
shows the curve for $K=1$ and $\rho=0.01$.}
\label{multi_compare}
\end{center}
\end{figure}

\Fref{multi_compare} shows the semi-log plot of the 
default probability $P_{N}(n)$ for $K=1,2,10,50,100$. 
We set the model  parameters as $p=\rho=0.03$,
 $N=100$, $\rho_{inter}=0.01$ and $p_{y}=0.5$.
The solid curve plots the data for $K=1$ and $\rho=0.03$ and the
dotted line shows the data for $K=1$ and $\rho=0.01$. As $K$ increases,
 the data for each $K$ departs from the solid  line. At $K=10$, the 
data almost shrinks on  the dotted line ($K=1$ and $\rho=0.01$).  
The data for $K=50$ almost coincide with those of $K=100$, whose $P_{N}(n)$
is given by the superposition of the two binomial distributions. 
This point is also a drawback of the present model.
If the glueing process works perfectly, these data should 
coincide with the homogeneous portfolio case $K=1$ and $\rho=0.01$.
However, this discrepancy is inherent property of the model.
Contrary to the discrepancy in   the two-sector case, the
conditions on the conditional default probabilities
$p^{k}_{n_{k},1}$ are different from those of the $K=1$ homogeneous 
portfolio. They could cause the difference in $P_{N}(n)$.

\begin{figure}[htbp]
\begin{center}
\includegraphics[width=12cm]{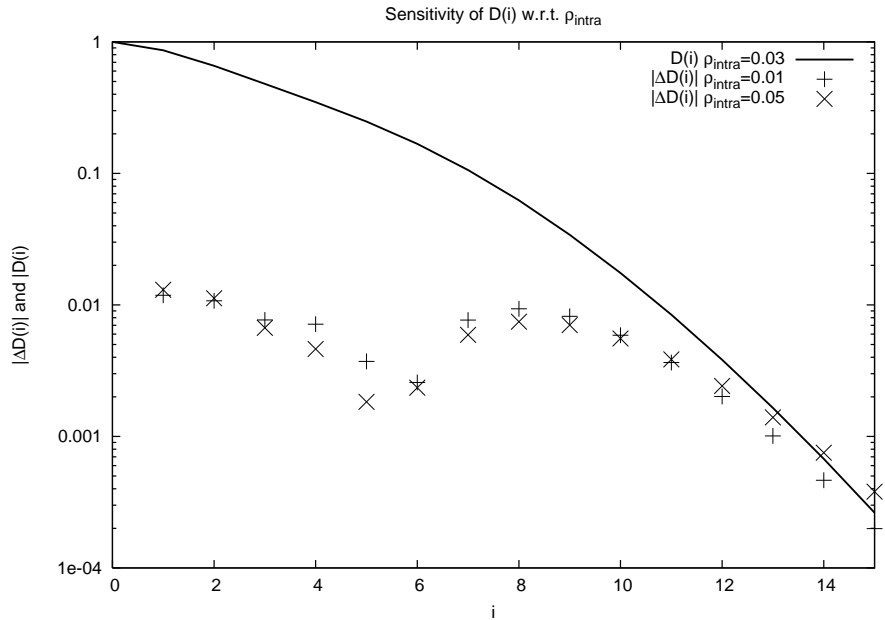}
\caption{   
Plot of $|\Delta D(i)|$ and $D(i)$ vs $i$.
$\Delta D(i)$ are between $\rho=0.01,0.05$ and $\rho=0.03$. 
$N=100$, $K=2,10,100$, $p=0.03$, $\rho_{inter}=0.01$ 
and $\lambda=0.3$. We set $p_{y}=0.5$.
The solid curve shows the data of $D(i)$ with $\rho=0.03$.}
\label{multi_compare_2}
\end{center}
\end{figure}

From the above discussions on $P_{N}(n)$ with different $K$,
we think that the inter-sector default correlation $\rho_{inter}$
is more important than 
the intra-sector default correlation $\rho$ in cases of a large $K$.
In the $K=10$ case, the $P_{N}(n)$ are roughly given by those of
 the homogeneous portfolio with $\rho=\rho_{inter}$.
If one estimates the implied values of $\rho$ and $\rho_{inter}$
from the premium (or $D(i)$) of the portfolio with large $K$,
this point is crucial. 
In \Fref{multi_compare_2}, we show $|\Delta D(i)|$
 for $K=10$ between  
$\rho=0.01,0.05$ and $\rho=0.03$. 
We set the model parameters as in the previous figure.
For comparison, we also plot $D(i)$ for $\rho=0.03$.
$D(i)$ represents the expected loss rate of the $i$-th tranche,
the magnitude of $|\Delta D(i)|$ is important when one estimates
the implied default correlation $\rho$ from the premium of the tranche.   
We see that $|\Delta D(i)|$ with small $i$ is small as compared with
$D(i)$. 
If we change $\rho$ from $\rho=0.03$, $D(i)$ does not change significantly.
It is difficult to obtain the implied values of $\rho$ from
the premium of the tranche with lower seniority.
In contrast with medium values of $i \simeq 13$, 
the magnitudes of $D(i)$ and $\Delta D(i)$ are almost comparable. 
$D(i)$ is sensitive to the change in $\rho$ and 
it is not difficult to derive the implied value of $\rho$.

\section{Implied Default Correlation}
\label{implied_c}
 
In the last section, as a concrete example, we try to
estimate the implied values of the default correlation
from the premium of a synthetic CDO. 
We treat  iTraxx-CJ (Series 2), which is an equally weighted portfolio
of 50 CDSs on Japanese companies. The standard attachment points 
and detachment points are $\{0,3\%\}$,$\{3\%,6\%\}$,$\{6\%,9\%\}$,$
\{9\%,12\%\}$ and $\{12\%,22\%\}$. Table \ref{premium} shows quotes 
on July 5,2005. The quote for the $\{0\%,3\%\}$ transhe shows the
upfront payment (as a percent of principal) that must be paid in
addition to 300 basis points per year. The other quotes for the other 
tranches are the annual payment rates in basis points per year. 
The index indicates the cost of entering into a CDS on all 50 companies
underlying the index. The recovery rate $R$ is $0.35$.

\begin{table}[hbtp]
\begin{center}
\caption{Quotes for iTraxx-CJ Tranches on July 5,  2005. 
Quotes for the $\{0,3\%\}$ tranche are the percent of the principal that
 must be paid up front in addition to 300 basis points per year.
Quotes for other tranches and the index are in basis points.
Source: Morgan Stanley Japan Securities Co. and Bloomberg
}
\vspace*{0.4cm}
\begin{tabular}{|l|c|c|c|c|c|c|} \hline
 & $\{0,3\%\}$ & $\{3\%,6\%\}$ & $\{6\%,9\%\}$  & $\{9\%,12\%\}$ &
$\{12\%,22\%\}$ & Index  \\
\hline
5-year Quotes & 15.75 & 113.25  & 42.0 & 30.5   & 15.5 & 24.55  \\
\hline 
\end{tabular}
\label{premium}
\end{center}
\end{table}

In order to get the implied default correlation for each 
tranche, 
it is necessary to
relate the loss distribution function $P_{N}(n)$ to the premiums. The
premiums are the present value of the expected cash flows. 
The calculation of this present value involves three terms \cite{Hull4}.
We denote by $P_{k}(a_{L},a_{H})$ the remaining notional for the
$\{a_{L},a_{H}\}$ tranche after 
$k$ defaults. It is given as
\begin{equation}
P_{k}(a_{L},a_{H})=
\left\{
\begin{array}{cc}
(a_{H}-a_{L})N & k< \lceil a_{L}N /(1-R) \rceil \\
a_{H}N-k(1-R)  & \lceil a_{L}N/(1-R) \rceil  \le 
 k  \le  \lceil
 a_{H}N/(1-R) \rceil  \\
0  &  k \ge \lceil a_{H}N/(1-R)) \rceil \\
\end{array}
\right.
\end{equation}
Here, $\lceil x \rceil$ means the smallest integer greater than $x$.
For simplicity,
we treat the 5-year as one period. The three terms are written as
\begin{eqnarray}
A=5.0 \times <P_{k}(a_{L},a_{H})>e^{-5.0r} \nonumber \\
B=2.5 \times ((a_{H}-a_{L})N-<P_{k}(a_{L},a_{H})>)e^{-r\frac{5}{2}}
 \nonumber \\
C=((a_{H}-a_{L})N-<P_{k}(a_{L},a_{H})>)e^{-r\frac{5}{2}}
\end{eqnarray}
 where $r$ is the risk-free rate of interest and we set $r=0.01$.
By considering the total value of the contract, one can see that 
the break even spread is given as $C/(A+B)$.

For the index, $a_{H}=1.0$ and $a_{L}=0.0$ and it is possible
to estimate the average default probability $p$. Instead, we use 
the CDS data for each company and estimate the average default 
probability $p$ and its  dispersion  
$\Delta P_{d}$ as
\[
p=1.8393 \% / \mbox{5-year}  
\hspace*{0.5cm}
\Delta P_{d}=1.131 \%  / \mbox{5-year}.
\]

With these parameters, the tranche correlations can be implied 
from the spreads quoted in the market for particular tranches.
These correlations are known as tranche correlations or compound correlations. 
As a pricing model, we use MCB, BBD and Gaussian
copula models. For MCB, we use the following candidates.
\begin{itemize}
\item  Original MCB model (MCB1). $N=50,K=1$ and $\lambda=0.0$. 
\item  Short tail MCB model (MCB2). $N=50,K=1$ and $\lambda=0.3$. 
\item  Short tail MCB model (MCB3). $N=50,K=1$ and
       $\lambda=0.6$. 

\item  Disordered MCB model (MCB4). $N=25+25$,$\lambda=0.3$,
$\rho_{xy}=\rho_{x}=\rho_{y}=\rho$ and
       $p_{x}=p+\Delta P_{d},p_{y}=p-\Delta P_{d}$. $N+M$ MCB model with
       inhomogeneous default probability.
\item  Two-sector MCB model (MCB5). $K=2$,$\rho_{x}=\rho_{y}=\rho,p_{x}=p_{y}=p$ and 
$\rho_{xy}=0.0$. Assets are categorized in $2$ sectors and $\rho_{inter}=0.0$.  
\end{itemize}

\begin{table}[hbtp]
\caption{Implied tranche correlation (\%) for 5-year iTraxx-CJ on July 5,2005.}
\vspace*{0.5cm}

\begin{center}
\begin{tabular}{|l|c|c|c|c|c|c|c|} \hline
Tranches & MCB 1 &MCB 2 &  MCB 3  & MCB 4&  MCB5 &BBD & Gaussian  \\ 
\hline
$\{0\%,3\%\}$   &  11.79  &  10.8  & 9.96 & 12.88  & 21.2  &  11.4   & 13.8  \\
$\{3\%,6\%\}$   &  1.27   &  1.18  & 1.13 & 1.36  & 2.45  &  1.26   & 1.35 \\
$\{6\%,9\%\}$   &  3.16   &  3.08  & 3.09 & 3.46  & 6.32  &  3.15   & 3.23  \\
$\{9\%,12\%\}$  &  6.16   &  5.95  & 5.90 & 6.65  & 12.15 &  6.11   & 6.31  \\
$\{12\%,22\%\}$ &  9.78   &  9.67  & 9.90 & 10.67  & 19.97 &  9.73   & 9.46  \\
\hline 
\end{tabular}
\label{implied}
\end{center}
\end{table}

Table \ref{implied} shows implied tranche correlations for the 5-year
quotes in Table \ref{premium}. 
We see a  ``Correlation
Smile'', which is a typical behavior of implied correlations 
across portfolio tranches \cite{Andersen}. 

We also find  that the implied correlations are different 
among the models. 
For MCB, models  with a larger $\lambda$
 have a  smaller correlation skew. The correlation
for $\{0\%,3\%\}$ decreases with $\lambda$ and other correlations
do not change significantly. If a probabilistic model describe the true default
distribution, there should not exist any correlation skew.
As a model approaches the true distribution, we can expect that the
skew decreases. MCB3  is more faithful to the true
default distribution. The skew of BBD is between MCB 1 and MCB 2, which
is reasonable because the profile of BBD is between MCB1 and MCB2 (
see \fref{Dist}).
 Gaussian copula's skew range is larger than
MCB 1, MCB 2, MCB 3 and BBD. 

About the effect of $\Delta P_{d}$, 
the implied correlations are considerably
different  between $\Delta
P_{d} \neq 0$ (MCB 4) and $\Delta P_{d}=0.0$ (MCB 2). 
In the estimation of the implied correlation, we cannot neglect the
fluctuation $\Delta P_{d}$. In particular, for the tranches 
$\{0\%,3\%\}$, the implied value is affected greatly by $\Delta P_{d}$.
As has been discussed in \cite{Finger2}, the increase in the 
dispersion of the default probability increases the loss in the equity tranche.
It is necessary to increase the implied correlation to match with the market
quote.

In the case $K=2$  (MCB5), the correlation level and correlation 
skew are very large.
This means that if assets are categorized in many sectors and
inter-sector correlation is very weak, the loss distribution 
accumulates around the origin $n=0$. 
In order to match with the market quote, a 
 large intra-sector correlation 
is necessary.

\begin{figure}[htbp]
\begin{center}
 \includegraphics[width=12cm]{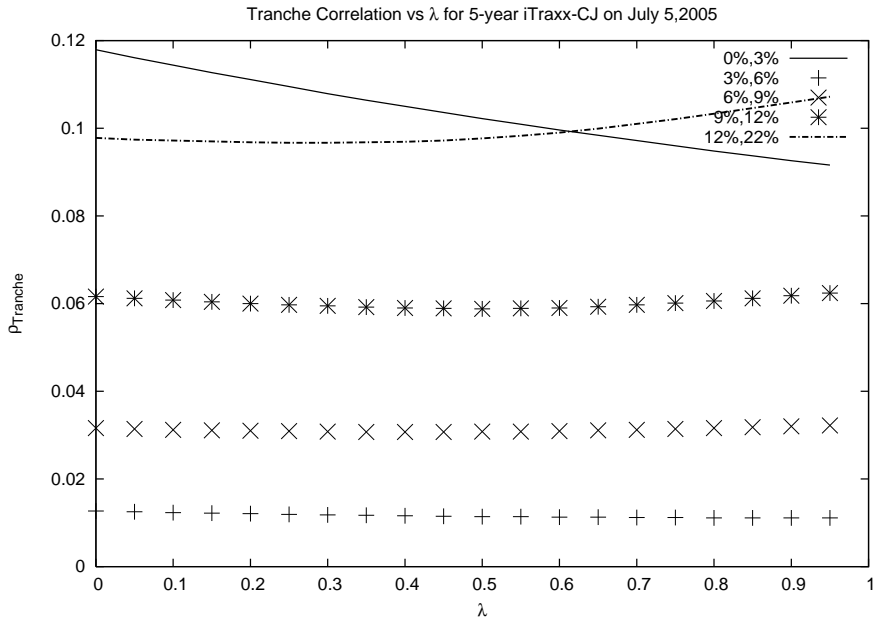}
\caption{Range pf implied tranche correlation for 5-year iTraxx-CJ on July 5, 2005
 vs $\lambda$.}
\label{Tranche}
\end{center}
\end{figure}

For the estimation of the model parameter $\lambda$, we think that the
resulting loss distribution should look closer to the implied 
loss distribution. That is, the range of tranche correlation skew should be
small. \Fref{Tranche} shows the tranche correlation vs $\lambda$.
As we increase $\lambda$, the skew range becomes small.
At $\lambda \simeq 0.61$, the range becomes minimal.
We should calibrate $\lambda$ to be $\lambda=0.61 \sim 0.62$.

\section{Concluding Remarks}
\label{conclusion}

In this paper, we generalize  Moody's correlated binomial
default distribution to the inhomogeneous portfolio cases.
As the inhomogeneity, we consider the non-uniformity in the 
default probability $p$ and in the default correlation $\rho$ and
$\rho_{inter}$. To treat the former case, we construct a coupled
 $N+M$ MCB model
and obtain the default probability function $P_{N+M}(n)$.
The inhomogeneity in $p$ causes  changes
in the expected loss rates of the tranches with lower seniority.

In order to treat the inhomogeneity 
in the default correlation, 
we construct a multi-sector MCB model by glueing multiple MCB 
models by an auxiliary
random variable $Y$. We cannot take out the joining lines  between the MCB
models, for small portfolio and small $\rho_{inter}$, the 
construction works well. For the inhomogeneity in $\rho$, we 
divide a homogeneous portfolio into  $K$ sectors. 
We set $\rho_{inter} <\rho=\rho_{intra}$ 
and see the effect of the increase in $K$
on $P_{N}(n)$. As the sector number $K$
increases, the inter-sector correlation $\rho_{inter}$ becomes 
more important  than the intra-sector default correlation $\rho$.
With large $K$, the default correlation is governed by $\rho_{inter}$
only.  The CDOs, whose assets are categorized in many sectors, 
$\rho_{inter}$ should be treated more carefully than $\rho=\rho_{intra}$.

In order to check the validity of the MCB model and our generalization
method, more careful treatment and calibration should be done.
We assume that $\rho_{n}$ decays exponentially with $n$.
With such a modification, the skew of the 
correlations diminishes, however, the skew remains significantly.  
Other models for $\rho_{n}$ should be considered. For this purpose, it is
necessary
to study the implied loss distribution directly.   
 Recently, Hull and White \cite{Hull3,Hull4}  developed 
a method to derive the implied loss distribution and to obtain
 the implied copula function from the market quotes of CDOs.
The authors proposed a calibration method for $\rho_{n}$
from the implied loss function \cite{Mori2}.
 By incorporating this information in the MCB model's framework,  
we may have a 
``Perfect'' correlated binomial default distribution model 
which reflects  market quotes completely. We think that 
 our  generalization method provides important information
based directly on the market quotes.

\section{Acknowledgement}

This research was partially supported by the Ministry of Education,
Science, Sports and Culture, Grant-in-Aid for 
Challenging Exploratory Research ,21654054, 2009.

\end{document}